\documentclass[onecolumn]{aastex631}

\usepackage{pifont}
\usepackage{mathtools}

\usepackage{graphicx}
\usepackage{amsmath}
\usepackage{natbib}
\usepackage{amssymb}
\usepackage{hyperref}
\usepackage{booktabs}
\usepackage{xcolor}
\usepackage{siunitx}
\usepackage{mathrsfs}

\newcommand{\pvnot}{\left(\Gamma\beta\right)_{\rm sh,0}}

\newcommand{\inj}[1]{\rm {inj}}
\newcommand{\etath}{z_\infty}
\newcommand{\ymax}{ y_{\rm t}}

\begin{document}
\title{Synchrotron Emission from Cooled Particle Distributions}

\correspondingauthor{Ross Ferguson}
\email{fergu734@umn.edu}

\author[0009-0004-9306-291X]{Ross Ferguson}
\affiliation{School of Physics and Astronomy, University of Minnesota, Minneapolis, MN 55455, USA}

\author[0000-0001-8405-2649]{Ben Margalit}
\affiliation{School of Physics and Astronomy, University of Minnesota, Minneapolis, MN 55455, USA}

\begin{abstract}

\end{abstract}

\begin{abstract}
Synchrotron emitting electrons can lose energy (`cool') through various processes including radiative losses (e.g., synchrotron or inverse-Compton cooling) and adiabatic expansion.
Such cooling will shift electrons in energy-space and therefore change the electron distribution function. This
in turn alters the nature of synchrotron emission and absorption from these electrons. In past literature these effects have typically been considered using either simplified one-zone frameworks, or using numerical methods as part of more accurate local modeling. In this work we extend the latter `local' treatment by deriving analytic expressions that are both accurate and more computationally efficient than previous numerical approaches. Considering two concrete cases of injected power-law and thermal electron distribution functions, we derive analytic fitting functions for the resulting emission and absorption coefficients including the effects of cooling. These fitting functions can be applied to synchrotron afterglow modeling from a variety of astrophysical sources, such as gamma-ray bursts (GRBs), luminous fast blue optical transients (LFBOTs), and jetted tidal disruption events (TDEs).
\end{abstract}


\section{Introduction}\label{intro}

Synchrotron emission from relativistic electrons occurs in a wide variety of astrophysical sources, including radio supernovae, gamma-ray bursts (GRBs), luminous fast blue optical transients, neutron star mergers, and tidal disruption events \citep[e.g.,][]{Weiler86,Chevalier98,Sari98,Burrows11,Berger14,Margutti19,Ho19,Nakar11,Kathirgamaraju2019}. In these contexts, synchrotron emission is assumed to originate in the downstream region of strong shock waves, with the shock providing a mechanism to accelerate electrons to relativistic velocities.

Modeling of synchrotron-emitting astrophysical sources typically falls into two general categories: one-zone or full-volume models. In global one-zone models \citep[e.g.,][]{Sari98,Margalit&Quataert21}, there is no spatial variation and electrons are injected and cool uniformly. Two advantages of one-zone models are that a detailed hydrodynamic history of the emitting fluid does not need to be specified and that the emission and absorption coefficients can be approximately parameterized analytically. By contrast, full-volume models \citep[e.g.,][]{Granot+99a,Granot+99b,ResslerLaskar17,Warren22,FM26} treat the three-dimensional structure of the emitting region (the shock and downstream fluid) self-consistently, and there can be considerable spatial variation in the fluid properties. Such models are more complicated and usually require numerical treatments, but they more accurately represent our conception of the underlying physics and often make substantively different predictions than one-zone models \citep{FM26}. Though convenient, the advantages of one-zone models in the absence of calibration by full-volume models make them prone to error. In full-volume models, the downstream electrons cool radiatively and adiabatically, altering the electron distribution and leading to highly spatially-dependent emission and absorption coefficients. The cooled radiation coefficients then depend on the post-shock hydrodynamics and cannot generally be solved analytically.
Previous approaches have therefore employed numerical methods to compute the cooled electron distribution function and associated emission and absorption coefficients at each point behind the shock \cite[e.g.,][]{GS02,ResslerLaskar17,Warren22}. This increases the computational cost of running such models. The aim of our current work is to circumvent this problem by deriving analytic expressions for the local electron distribution function and its resulting synchrotron emission and absorption coefficients, accounting for adiabatic and synchrotron cooling.

Past modeling efforts have typically assumed that the synchrotron-emitting electrons follow a power-law distribution in energy, motivated by studies of first-order Fermi acceleration \citep[]{Bell,BE87,BO78}. Recent particle-in-cell (PIC) simulations support the idea that a power-law distribution develops in the downstream fluid of shocks, but indicate the presence of a thermal distribution in addition \citep[e.g.,][]{Spitkovsky08,Sironi13,Park11,Crumley19,Jikei25}. Thermal synchrotron emission has been invoked in several studies \citep[e.g.,][]{Ozel, GianniosSpitkovsky09, ResslerLaskar17,Warren22,Margalit&Quataert21}.
Considering both power-law and thermal injected electron distributions, we provide analytic fitting functions for the radiation coefficients. These fitting functions can be used in any context where an electron distribution is impulsively injected at a single time and subsequently cools.  A primary application of these fitting functions is to full-volume models of synchrotron-emitting shocks. These fitting functions are a considerable improvement over computing the full numerical integrals defining the radiation coefficients, and allow a for a quick, consistent treatment of cooling in synchrotron spectra. 

We begin in \S\ref{sec:distributions} by considering an arbitrary,  impulsively injected electron distribution subject to radiative and adiabatic cooling. We then specialize to two cases, a power-law and a relativistic Maxwellian, and analytically calculate the respective cooled distribution functions. In \S\ref{sec:pl_dist}, we examine the emission and absorption coefficients associated with the cooled power-law distribution and provide fitting functions for each case. Analogous fitting functions for the thermal emission and absorption coefficients are given in \S\ref{sec:th_dist}. In \S\ref{sec:afterglow}, we apply the power-law fitting functions to a full-volume model of GRB afterglows and compare to the results of \cite{GS02}. We conclude in \S\ref{sec:conclusion}.

\section{General Description of Downstream Cooling}\label{sec:distributions}

In the conventional picture of astrophysical shock acceleration, interaction between swept-up electrons and the shock front leads to the injection of a power-law distribution in energy, $(\partial n /\partial \gamma)_{\rm inj, pl} \,\propto \gamma^{-p}$. As mentioned in the previous section, PIC simulations indicate that electrons are accelerated at the shock to both a power-law distribution and a thermal distribution. For now, we consider the evolution of a generic distribution function, and return to these particular cases later.

When a particle distribution $(\partial n /\partial \gamma)_{\rm inj}$ is impulsively injected at the shock front, the particles cool as they are advected downstream by the fluid. We assume that the injection of electrons occurs only at the shock front, so that the total number of electrons $N_e$ in a fluid element is constant in time. Then, with $n_e$ denoting the electron number density,
\begin{equation}\label{eq:electron_conservation}
    \frac{d}{dt} \int d\gamma \,\,\frac{\partial N_e}{\partial\gamma} = N_e\frac{d}{dt} \int d\gamma \,\,\ \frac{1}{n_e} \frac{\partial n_e}{\partial\gamma} = 0.
\end{equation}
Thus, at any time $t>t_{\rm inj}$ after the time at which electrons are injected ($t_{\rm inj}$), we may take $d\gamma \,(\partial n_e/\partial\gamma)/n_e$ to be constant in time to obtain \citep{GS02,ResslerLaskar17,Warren22}
\begin{equation}\label{eq:distribution_evolution}
    \left(\frac{\partial n_e}{\partial \gamma}\right) = \frac{n_{e}}{n_{e, \rm inj}} \,\frac{d\gamma_{ \rm inj}}{d\gamma} \,\left(\frac{\partial n_e}{\partial \gamma}\right)_{\rm inj},
\end{equation}
where quantities without the subscript `$\rm inj$' implicitly refer to quantities evaluated at time $t$. The distribution function $(\partial n_e/\partial \gamma)_{\rm inj}$ is set by the initial injection at the shock front, while the dilution of the number density $n_{e}/n_{e, \rm inj}$ is determined by the hydrodynamics. The relation between $\gamma_{ \rm inj}$ and $\gamma$ must be fixed by examining the cooling processes in detail. In other words, we must solve for the energy of an electron as a function of time.

\subsection{Electron Cooling}
We consider two sources of cooling: radiative cooling, which occurs in the present context by the emission of synchrotron radiation, and adiabatic cooling, which is caused by $PdV$ work done as the fluid element expands. 
Accounting for these two processes, 
the instantaneous cooling rate for an electron with Lorentz factor $\gamma$ is \citep[e.g.,][]{GS02, Zhang19, AguilarRuiz25}
\begin{equation}\label{eq:cooling_ODE}
    \frac{d\gamma}{dt} = -\frac{\sigma_T B^2}{6\pi m_e c}\gamma^2 +(\hat{\gamma}-1) \frac{d \log n_e}{dt} \gamma = -\frac{\gamma^2}{t_B}-\frac{\gamma}{t_{\rm ad}},
\end{equation}
where $\sigma_T$ is the Thomson cross-section, $B$ is the local magnetic field strength, $n_e$ is the local electron density, and $\hat{\gamma}$ is the adiabatic index of the electrons. The time $t$ is the time since injection; in the case of cooling in the downstream region of a shock, $t$ is measured in the fluid rest frame. The first term corresponds to synchrotron cooling, 
and we have defined a characteristic timescale $t_B$ such that $t^{-1}_B = \sigma_T B^2 /6\pi m_e c$. 
This is related to the synchrotron cooling time $\sim t_B / \gamma$ over which an electron with Lorentz factor $\gamma$ radiates an order unity fraction of its energy.
The second term in Equation~(\ref{eq:cooling_ODE})
corresponds to adiabatic cooling and is associated with an inverse timescale $t^{-1}_{\rm ad} = -(\hat{\gamma}-1) d \log n/dt$. Note that enhanced cooling due to inverse-Compton scattering can easily be taken into account by dividing $t_B$ by $1+Y$, where the Compton Y-parameter $Y=P_{\rm IC}/P_{\rm syn}$ is the ratio of inverse Compton to synchrotron power \citep{Zhang19}. For simplicity, we set $Y=0$ in the following.

Since the timescales $t_B$ and $t_{\rm ad}$ are independent of the electron Lorentz factor, Equation~(\ref{eq:cooling_ODE}) is a form of Bernoulli's equation and may be formally solved for the Lorentz factor of a given electron as a function of time since injection.\footnote{In the application to a relativistic fluid, the time since injection should be evaluated in the fluid rest frame.} 
The integration constant is set by the initial condition $t=t_{\rm inj}$, for which we define $\gamma (t_{\rm inj})= \gamma_{ \rm inj}$. The time evolution of an electron's Lorentz factor $\gamma(t)$ can therefore be written as
\begin{equation}\label{eq:gamma_loc}
    \gamma(t) = \frac{\mathscr{G}^{-1}(t)}{\gamma^{-1}_{ \rm inj} + \mathscr{F}(t)} = 
    \frac{\gamma_{ \rm inj}}{\mathscr{G}(t) + \gamma_{ \rm inj}/\gamma_{\infty}},
\end{equation}
where we have defined the dimensionless functions
\begin{equation}\label{eq:G}
  \mathscr{G}(t) =\exp\left[\displaystyle\int_{t_{\rm inj}}^{t} dt_1 \,t^{-1}_{\rm ad}\right],
\end{equation}
\begin{equation}\label{eq:F}
  \mathscr{F}(t) =\displaystyle\int_{t_{\rm inj}}^{t} dt_1 \,\, t^{-1}_B \,\,\exp\left[- \displaystyle\int_{t_{\rm inj}}^{t_1} dt_2 \,\,t^{-1}_{\rm ad}\right] = \displaystyle\int_{t_{\rm inj}}^{t} dt_1 \,\, t^{-1}_B \,\, \mathscr{G}^{-1}(t_1).
\end{equation}
In Equation~(\ref{eq:gamma_loc}),
$\gamma_{\infty}$ is defined to be the Lorentz factor that an electron injected with $\gamma_{\rm inj} = \infty$ has at a time $t>t_{\rm inj}$,
\begin{equation}\label{eq:gamma_inf}
    \gamma_\infty(t) \equiv \frac{1}{\mathscr{G}(t)\mathscr{F}(t)}.
\end{equation}
Inverting Equation~(\ref{eq:gamma_loc}), the injection Lorentz factor of an electron with $\gamma$ at time $t$ and the derivative needed to calculate the evolution of the distribution function are
\begin{equation}\label{eq:gamma_inj}
    \gamma_{ \rm inj} = \mathscr{G}(t)
    \frac{\gamma}{1- \gamma/\gamma_{\infty}}, \hspace{50pt}
    \frac{d\gamma_{ \rm inj}}{d\gamma} =
    \frac{\mathscr{G}(t)}{(1- \gamma/\gamma_{\infty})^2}.
\end{equation}
The equations presented in this subsection are general to any hydrodynamic profile. The impact of the hydrodynamics is to specify $\mathscr{F}$ and $\mathscr{G}$ as a function of time for a given fluid element. In general, these functions must be solved numerically. However, certain cases, such as a Blandford-McKee hydrodynamic solution, permit an analytic solution \citep{GS02}. For completeness, we review these results in detail in Appendix~\ref{appendix:Blandford-McKee}. 

\subsection{Cooled Distribution Functions}

In this paper we discuss the radiation coefficients corresponding to two different injected distribution functions: a simple power-law and a relativistic Maxwellian. These distributions take the form, in the absence of cooling,
\begin{equation}
    \left(\frac{\partial n_e}{\partial \gamma}\right)_{\rm pl,\, inj} =K \,\gamma^{-p},
\end{equation}
\begin{equation}\label{eq:thermal}
      \left(\frac{\partial n_e}{\partial \gamma}\right)_{\rm th, \,\rm inj} = L\,\sqrt{1-1/\gamma^2} \,\,\frac{\gamma^2}{2\Theta^3}\,\,e^{-\gamma/\Theta},
\end{equation}
where $\Theta = kT_e/m_e c^2$ is the dimensionless electron temperature and the factors $K$ and $L$ denote the overall normalizations of the respective distributions. We write the total distribution function of the electrons as the sum of the thermal and power-law contributions, treating each contribution separately. The power-law distribution is assumed to extend over Lorentz factors $\gamma_1\leq\gamma\leq\gamma_2$, while the thermal distribution has the domain $1\leq\gamma\leq\gamma_3$. In the formalism of \cite{Margalit&Quataert21}, $\gamma_2 = \gamma_3 = \gamma_\infty$ and $\gamma_1 = 1+a(\Theta)\Theta$ is chosen such that $\gamma_1$ is the mean Lorentz factor of the thermal distribution, where $a(\Theta) \simeq (6+15\Theta)/(4+5\Theta)$. 
This choice of Lorentz factors leads to a discontinuity in the hybrid distribution function, since $(\partial n/\partial \gamma)_{\rm th} + (\partial n/\partial \gamma)_{\rm pl} > (\partial n/\partial \gamma)_{\rm th}$ at $\gamma_1$. In the shocks of interest, the number and energy fraction contained in power-law electrons is small, and this unphysical discontinuity does not greatly affect the resulting emission. In fact, it is inherent to typical treatments of power-law distributions in the literature due to the truncation at Lorentz factor $\gamma_1$.\footnote{To avoid this issue, other authors have chosen a different parameterization for the power-law-thermal hybrid distribution where $\gamma_3 = \gamma_1$ and the overall normalizations are changed such that the hybrid distribution function is continuous \citep[e.g.,][]{Yuan03,Giannios09,ResslerLaskar17,Warren22}. Importantly, this choice makes it harder to find fitting functions, as the thermal and power-law distributions are no longer independent of one another and cannot be characterized separately (this choice introduces an additional free parameter). Since any potential corrections needed to deal with the discontinuity of the combined distribution are sub-leading at high $\Theta$ and $\gamma_1$, we neglect them in this work.}

\begin{figure*}
 \centering
  \includegraphics[width=0.9\textwidth]{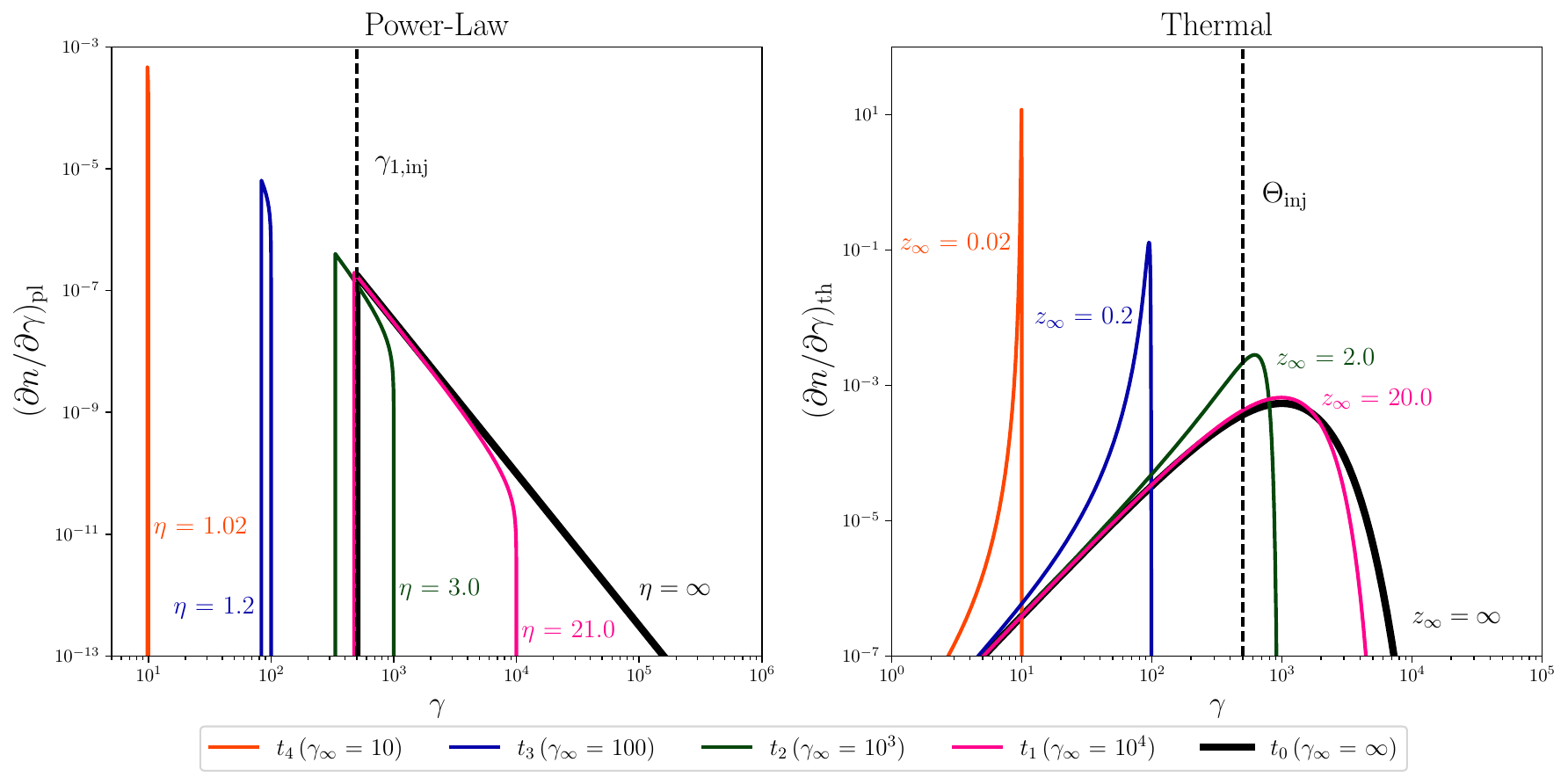}
  \caption{Cooled power-law (left) and thermal (right) electron distribution functions (Equations~\ref{eq:pl_dist_cooled},\ref{eq:therm_dist_cooled}). The blue curves show the initial distributions, which extend to Lorentz factors of $\infty$. As time increases, $\gamma_\infty$ falls (Equation~\ref{eq:gamma_inf}), changing both the form of the distribution and the location of the high-frequency cutoff. The dashed black lines show the injected values of the minimal power-law Lorentz factor $\gamma_{1, \rm inj}$ and temperature $\Theta_{\rm inj}$, respectively. As time increases, both distributions become strongly peaked, resembling a Dirac delta function in energy. For each curve, the values of the parameters $\eta = \gamma_\infty/\gamma_1$ and $\etath = \gamma_\infty \mathscr{G}/\Theta_{\rm inj}$, discussed in detail in later sections, are labeled for convenience. Note that adiabatic cooling is ignored ($\mathscr{G}=1$) and the pre-factors in each distribution (which in general depend on time and the assumed hydrodynamics) are set to 1.}
  \label{fig:distributions}
  \end{figure*}

To find the form of the cooled distribution functions, we substitute Equation~(\ref{eq:gamma_inj}) into Equation~(\ref{eq:distribution_evolution}) to obtain
\begin{equation}\label{eq:pl_dist_cooled}
    \left(\frac{\partial n_e}{\partial \gamma}\right)_{\rm pl} = K_{\rm inj}\,\frac{n_e}{n_{e, \rm inj}} \, \mathscr{G}^{1-p} \,\,
    \gamma^{-p} \left(1- \frac{\gamma}{\gamma_{\infty}}\right)^{p-2},
\end{equation}
\begin{equation}\label{eq:therm_dist_cooled}
    \left(\frac{\partial n_e}{\partial \gamma}\right)_{\rm th} = L_{\rm inj} \,\frac{n_e}{n_{e, \rm inj}} \,\frac{\gamma^2}{2\Theta_{\rm inj}^3}\, \frac{\mathscr{G}^3}{(1- \gamma/\gamma_{\infty})^4}\,\exp \left[ -\frac{\mathscr{G}}{\Theta_{\rm inj}} \frac{\gamma}{1- \gamma/\gamma_{\infty}}\right] \times \sqrt{1-\frac{1}{\mathscr{G}^2}\frac{(1- \gamma/\gamma_{\infty})^2}{\gamma^2}}.
\end{equation}
These are the full forms for the cooled downstream distributions. Schematic versions (ignoring adiabatic cooling and setting the prefactors to 1) are shown in Figure~\ref{fig:distributions}. The full forms differ from the injected distributions by the prefactors $\mathscr{G}$ and $n_{e}$ (due to adiabatic cooling and dynamical changes to the downstream fluid); the change to $\gamma_1$, $\gamma_2$, and $\gamma_3$ due to cooling; and the cutoff factors proportional to $1- \gamma/\gamma_{\infty}$. It is these differences which primarily concern us in the following sections, where we derive fitting functions for the emission and absorption coefficients arising from these distributions.

The cooled power-law distribution deserves special attention. First, in this formalism we cannot have $p<2$ because $(1-\gamma/\gamma_\infty)^{p-2}$ diverges as $\gamma\to\gamma_\infty$. This divergence is related to the fact that for $p<2$ and $\gamma_{2, \rm inj}=\infty$, the kinetic energy of the distribution is infinite. Calculating emission and absorption coefficients when $p<2$ thus requires integrating from $\gamma_1$ to $\gamma_{2} <\gamma_\infty$, with no changes to Equation~(\ref{eq:pl_dist_cooled}). The $(1-\gamma/\gamma_\infty)^{p-2}$ term causes a pile-up of electrons at the maximal Lorentz factor $\gamma_2$, leading to behavior qualitatively different to $p>2$. The primary application of interest in this work is first-order Fermi acceleration with $p>2$, so we consider only the range $2<p \lesssim 5$ below (the upper limit is an arbitrary choice defining the range in which we have validated that our fitting functions are accurate). Second, the range of power-law Lorentz factors $\eta \equiv \gamma_\infty/\gamma_1$ changes over time according to
\begin{equation}\label{eq:eta}
    \eta = 1 + \frac{1}{\gamma_{1, \rm inj} \mathscr{F} },
\end{equation}
which is always greater than or equal to 1. In the Blandford-McKee solution,  $\mathscr{F} \to \rm const.$ as the self-similar coordinate $\chi$ goes to $\chi\to \infty$  (equivalent to $t\to\infty$; see Appendix~\ref{appendix:Blandford-McKee}), and $\eta$ asymptotes to a value $\eta_{\rm f}>1$ (Appendix~\ref{appendix:Blandford-McKee}). In more general cases, $\mathscr{F}$ may diverge as $t\to\infty$, in which case $\eta\to1$. This issue can be traced back to the fact that in this limit we can have non-physical values of $\gamma_1<1$ and $\gamma_2<1$ --- that is, this formalism ignores any non-relativistic corrections to the emission pattern. Once $\gamma_1\lesssim2$, we should instead consider non-relativistic cyclotron radiation. Non-relativistic corrections are likely only relevant at late times in the evolution of the fluid and are
unlikely to be important in the cases of interest. We therefore ignore this limitation in our present work.

Given the dependence of $\eta$ on $\mathscr{F}$, it may be possible for the power-law distribution to cool significantly enough that $\eta -1 \ll 1$ while $\gamma_1>1$, so that the distribution function is approximately a Dirac delta function $\propto \delta(\gamma-\gamma_1)$ (for example, see the red curve corresponding to $\gamma_\infty=10$ in Figure~\ref{fig:distributions}). In this case, the emissivity is suppressed by a factor $\approx(1-\sqrt{\gamma_1/\gamma_2})^{p-2} \ll1$. Since this can occur in the physically interesting case where non-relativistic corrections are unimportant ($\gamma_1 \gg 1$), we include such effects in the power-law fitting functions below. Note that radiative synchrotron cooling is known to generate such Dirac-delta ring distributions \citep[e.g., ][]{Bilbao23,Bilbao24}, which can lead to the development of plasma instabilities and coherent radiation. The resulting effects on the emission properties are beyond the scope of this work.

\section{Fitting Functions: Power-Law Distribution}\label{sec:pl_dist}

In this section, we provide fitting functions for the power-law emission and absorption coefficients at an arbitrary point downstream, including the effects of cooling.

\subsection{Emissivity: \texorpdfstring{$j_{\nu, pl}$}{jnu pl}}
The general pitch-angle-averaged emissivity is \citep{RBL}

\begin{equation}\label{eq:jpl}
    j_{\nu, \rm pl} = \frac{1 }{4\pi } \int_{\gamma_1}^{\gamma_2} d\gamma \left(\frac{\partial n}{\partial \gamma}\right)_{\rm pl} \tilde{P}_e 
    = \frac{\sqrt{3}e^3 B }{4\pi m_e c^2} \int_{\gamma_1}^{\gamma_2} d\gamma \left(\frac{\partial n}{\partial \gamma}\right)_{\rm pl} \tilde{F}(x),
\end{equation}
where $\tilde{P}_e$ and $\tilde{F}$ are the pitch-angle-averaged power and synchrotron function, respectively (Appendix~\ref{appendix:synchrotron_function}). In this and the following equations, the variable $x$ is defined as
\begin{equation}
\label{eq:x}
    x \equiv \frac{4\pi m_e c \nu}{3eB\gamma^2} = \frac{\nu}{\nu_0\gamma^2},
\end{equation}
with the non-relativistic characteristic frequency $\nu_0 = 3eB/4\pi m_ec = 3\nu_B/2$ related to the electron gyrofrequency $\nu_B$. Physically, $x$ is  the ratio of the observed frequency to $\gamma_1^2\nu_0$, the characteristic synchrotron frequency of an electron with Lorentz factor $\gamma_1$. In terms of $x$, the emissivity integral runs from $x_\infty = \nu/\nu_2$ to $x_1 = \nu/\nu_1$, where $\nu_1=\gamma_1^2 \nu_0$ and $\nu_2=\gamma_2^2 \nu_0$ (recall the assumption $\gamma_2 = \gamma_\infty$, so that $\nu_2$ corresponds to $x_\infty$). Inserting the cooled power-law distribution function (Equation~\ref{eq:pl_dist_cooled}) into Equation~(\ref{eq:jpl}) gives
\begin{equation}
    j_{\nu, \rm pl} = \frac{\sqrt{3}e^3 B }{4\pi m_e c^2} K_{\rm inj}\,\frac{n_e}{n_{e, \rm inj}} \, \mathscr{G}^{1-p}\int_{\gamma_1}^{\gamma_2} d\gamma \,\, \tilde{F}(x) \,\gamma^{-p} \,\left(1- \frac{\gamma}{\gamma_{\infty}}\right)^{p-2}.
\end{equation}
Changing the integration variable to $x$, the emissivity takes the form
\begin{equation}\label{eq:emissivity}
    j_{\nu, \rm pl} = \frac{\sqrt{3}e^3 B }{8\pi m_e c^2} K_{\rm inj}\,\frac{n_e}{n_{e, \rm inj}} \, \mathscr{G}^{1-p}  
    \left(\frac{\nu}{\nu_0}\right)^{(1-p)/2}
    J_{\rm pl}(p, x_1, x_\infty),
\end{equation}
with a dimensionless function $J_{\rm pl}(p, x_1, x_\infty)$ encoding the part of the emissivity that requires a fitting function,
\begin{equation}\label{eq:J_pl}
    J_{\rm pl}(p, x_1, x_\infty) \equiv 
    \int_{x_\infty}^{x_1} dx \,\,\tilde{F}(x)\, x^{(p-3)/2} \,\left(1-\sqrt{x_\infty/x}\right)^{p-2}.
\end{equation}
The integration runs from $x_\infty = x(\gamma_2)$ to $x_1 = x(\gamma_1)$, and we have made the choice $\gamma_2=\gamma_\infty$, 
corresponding to an injected power-law distribution that extends to a Lorentz factor of $\infty$. In practice, this is equivalent to a finite maximum injected Lorentz factor so long as $\gamma_{2,{\rm inj}} \gg \gamma_{1,{\rm inj}}$ and $p>2$. For $p<2$, assuming $\gamma_{2,{\rm inj}}=\infty$ leads to an infinite electron kinetic energy. As mentioned above, assuming $p<2$ and a finite $\gamma_{2, \rm inj}$  leads to qualitatively different behavior, and we neglect this possibility in deriving analytic fitting functions below.

The function $J_{\rm pl}$ is similar to that considered by \cite{FO14}, with the added complication of the $1-\sqrt{x_\infty/x}$ factor. As in the case without this factor, there are three main regimes: low frequencies ($x_\infty<x_1\ll1$), where $j_{\nu, \rm pl} \sim \nu^{1/3}$; intermediate frequencies ($x_\infty\lesssim 1 \lesssim x_1$) with $j_\nu\sim \nu^{(1-p)/2}$; and high frequencies ($x_\infty\gg 1$), where the emissivity is exponentially cut off as $j_\nu\propto e^{-x_\infty} = e^{-\nu/\nu_0\gamma_2^2}$. The first two regimes are well understood; it is the presence of the cutoff and the $1-\sqrt{x_\infty/x}$ factor---which encapsulate the effects of cooling---which are novel in our derivation of analytic expressions for the radiation coefficients. In terms of $x_1$ and $\eta$, the different frequency regimes are $0 < x_1 < 1$ (low frequencies), $1 \leq x_1 < \eta^2$ (intermediate frequencies), and $\eta^2 \leq x_1 < \infty$. The key cutoff frequencies are $x_1 = 1$ and $x_1 = \eta^2$ (i.e., $x_\infty=1$).

\begin{table}
\centering
\begin{tabular}{@{}lccc@{}}
\toprule
     & Asymptotic Limit of $J_{\rm pl}$ 
      & Asymptotic Limit of $A_{\rm pl}$   \\ 
\midrule
 \textbf{Low Frequencies ($x_1\ll1$)}\\

    All $\eta$  &    $\Omega_p(\eta, x_\infty)$: Eq.~\ref{eq:Omega} &   $\chi_p(\eta, x_\infty)$: Eq.~\ref{eq:chi}       \\
  \midrule
     \textbf{Intermediate Frequencies ($1 <x_1 < \eta^2$)}\\
    $\eta-1\ll1$    &  $\Omega_p(\eta, x_\infty)$: Eq.~\ref{eq:Omega}  &  $\chi_p(\eta, x_\infty)$: Eq.~\ref{eq:chi}   \\
    $\eta\gg1$   &   $A_1(p)$: Eq.~\ref{eq:Jpl_inter}    & $B_1(p)$: Eq.~\ref{eq:B1}  \\
  \midrule
   \textbf{High Frequencies ($x_1\gg\eta^2$)}\\
    $\eta-1\ll1$    &  $\Omega_p(\eta, x_\infty)$: Eq.~\ref{eq:Omega}  &   $\chi_p(\eta, x_\infty)$: Eq.~\ref{eq:chi}     \\
    $\eta\gg1$   &   $\psi_p(\frac{p-3}{2}, x_\infty)$: Eq.~\ref{eq:psi_p}  &  $\psi_p(\frac{p}{2}, x_\infty)$: Eq.~\ref{eq:A_pl_high} \\
  \bottomrule                          
\end{tabular}
\caption{Summary of the asymptotic limits used to calculate $J_{\rm pl}$ and  $A_{\rm pl}$ in different limits of frequency ($x_1$) and energy range ($\eta$) and in which equations they are defined. For each frequency range, there are two asymptotic values of $\eta$ which generally have different limits. The functions $\Omega_p$ and $\chi_p$ each show up in several rows, a duplication due to the fact that when $\eta-1\ll1$, the distribution becomes a Dirac delta in energy and a single function applies at all frequencies. Note that at intermediate frequencies for $\eta-1 \ll 1$, there is no true asymptotic limit, since the intermediate-frequency regime is very small and acts only as a transition region between high and low frequencies. Nevertheless, the Dirac-delta distribution functions work in this regime and are labeled in the table.}
\label{table:Jpl_functions}
\end{table}

Following the approach of \cite{FO14}, we write the fitting functions in terms of $p$, $x_1$, and the range of Lorentz factors $\eta = \sqrt{x_1/x_\infty} = \gamma_\infty/ \gamma_1$. To accurately capture the behavior of the function $J_{\rm pl}$ at all frequencies, our strategy will be to find a fitting function $\Psi_p( \eta, x_1)$ fusing the intermediate and high frequency regimes which can then be joined to a low-frequency limit $\Omega_p( \eta, x_1)$. In the following subsections, we describe the various limits and the final fitting function joining each of them limits together. Table~\ref{table:Jpl_functions} provides a summary of where different limits of $J_{\rm pl}$ can be found.

\subsubsection{Low Frequencies \texorpdfstring{: $x_1\ll1$}{}}\label{sec:omega_pl}
At low frequencies, the integral giving $J_{\rm pl}$ may be solved exactly using
the low-frequency limit of the synchrotron function,
 $\tilde{F}(x \ll 1) \approx \tilde{F}_0 x^{1/3}$ (see Appendix~\ref{appendix:synchrotron_function}). Defining a new integration variable $u = \sqrt{x_\infty/x}$, we have
\begin{equation}
\begin{split}\label{eq:Omega}
     J_{\rm pl}(p, x_1\ll1,  x_\infty\ll1)  &  = 2 \tilde{F}_0\, x_\infty^{(3p-7)/6}\int_{\eta}^{1} du \,\,\, u^{-(3p+2)/3} \left( 1-u\right)^{p-2} \\
    & =     
    \frac{6}{3p-1} x_\infty^{(p-1)/2} \tilde{F}(x_\infty) 
    % \begin{aligned} 
    \left\{ \eta^{p-1/3} \,_2F_1 (2-p, \frac{1}{3}-p, \frac{4}{3}-p, \eta^{-1})   
        - \frac{\Gamma(p-1)\Gamma(\frac{4}{3}-p)}{\Gamma(-\frac{2}{3})}\right\} \\
    & \equiv \Omega_p(\eta, x_\infty),
    % \end{aligned} 
\end{split}
\end{equation}
where $ _2 F_1$ denotes the hypergeometric function. In the third line, we make use of the identities $\int du \,u^b (1-u)^a = u^{b+1} \,_2F_1 (-a,b+1,b+2,x)/(b+1) + \rm const.$ and $_2F_1 (-a,b+1,b+2,1) = \Gamma(a+1)\Gamma(b+2)/\Gamma(a+b+2)$. 

For $\eta\gg1$, the second term of Equation~(\ref{eq:Omega}) is negligible and the first term converges to $_2F_1 (-a,b+1,b+2,0) = 1$. After expanding the synchrotron function for small $x_\infty = x_1/\eta^2$, we are left with\footnote{This limit can be also be found by setting $x_\infty =0$ in Equation~(\ref{eq:J_pl}) and taking $x \ll 1$.}
\begin{equation}\label{eq:Omega_high_eta}
    J_{\rm pl} (p, \eta\gg1, x_\infty\ll1)  
    \approx \frac{6}{3p-1} \tilde{F}(x_1)\, x_1^{(3p-1)/6} 
   \underset{x_1\ll1}\approx \frac{6}{3p-1} \tilde{F}_0 \, x_1^{(3p-1)/6}.
\end{equation}
Conversely, in the limit $\eta-1\ll1$, cooling is so efficient that the injected electrons have all cooled to a Lorentz factor very close to $\gamma_1 \sim \gamma_2 = \gamma_\infty$. The distribution function is sharply peaked at a single Lorentz factor and is approximately proportional to $\delta(\gamma-\gamma_\infty) \propto\delta(x-x_\infty)$.  $J_{\rm pl}$ is then nearly proportional to the synchrotron function $\tilde{F}(x_\infty)$, taking the same form as the emissivity of a single electron but with a normalization given by Equation~(\ref{eq:Omega}). Equation~(\ref{eq:Omega}) is thus valid at all frequencies in the limit $\eta\to1$, provided we write it in terms of $\tilde{F}(x_\infty)$ instead of $\tilde{F}_0 x_\infty^{1/3}$.

To obtain a small-$\eta$ limit for $J_{\rm pl}$ one can either expand $_2 F_1(2-p,1/3-p,4/3-p,\eta)$ or directly expand the integrand of Equation~(\ref{eq:Omega_high_eta}) about $\eta-1 \ll 1$ to obtain\footnote{For very small $\eta-1 \lesssim 10^{-3}$, the general definition of $\Omega_p$ using the hypergeometric function fails numerically. The correct form of $\Omega_p$ in this case is given by Equation~(\ref{eq:Omega_low_eta}).}
\begin{equation}\label{eq:Omega_low_eta}
     J_{\rm pl} (p,\,  \eta-1\ll1, x_\infty\ll1)   
    \approx \,\frac{2}{p-1} x_1^{(p-1)/2} \tilde{F}(x_1) (\eta-1)^{p-1}.
\end{equation}

\subsubsection{Intermediate Frequencies \texorpdfstring{: $1 < x_1 < \eta^2$}{}}\label{sec:A_1}

For intermediate frequencies, $J_{\rm pl}$ is a constant. The value of this constant may be calculated using the definition of pitch-angle averaging in Appendix~\ref{appendix:synchrotron_function}. In this regime, we approximate $x_\infty\simeq 0$ and $x_1\gg1 $, which occurs when the observed frequency $\nu$ is much less than the frequency $ \nu_2$ corresponding to the maximal Lorentz factor and much greater than the frequency $\nu_1$ corresponding to the minimal Lorentz factor. Since 
\begin{flalign}\label{eq:Jpl_inter}
    J_{\rm pl}(p, x_1\gg1, x_\infty\ll1)  & = \int_0^{\pi/2} d\alpha \,\,\sin^2{\alpha} \,\int_0^{x_1} dx \,\,F(x/\sin{\alpha}) \,x^{(p-3)/2} & \nonumber \\
    & \simeq \int_0^{\pi/2} d\alpha \,\,\sin^{\frac{p+3}{2}}{\alpha} \, \int_0^{\infty} dz \,\,F(z) \,z^{(p-3)/2} & \nonumber \\
    & \simeq \frac{\sqrt{\pi} \,2^{\frac{p+3}{2}}}{(p+1)(p+3)} \frac{\Gamma(\frac{p}{4} + \frac{5}{4})\Gamma(\frac{p}{4} + \frac{19}{12})\Gamma(\frac{p}{4} - \frac{1}{12})}{\Gamma(\frac{p}{4} + \frac{3}{4})} \equiv A_1(p) .
\end{flalign}
In the second line, we change variables to $z = x/\sin{\alpha}$ and use $x_1\gg1$ to approximate the upper bound as $\infty$. The integrals may be evaluated using Equation 6.35a in \cite{RBL}. We have tacitly assumed $\eta\gg1$ in this subsection and the next, allowing us to ignore the finite value of $x_\infty$. The intermediate-frequency region extends over a range of $x$ equal to the value of $\eta$, so for $\eta -1 \sim 1$ the error incurred by assuming $x_\infty=0$ lasts for only a small range of frequencies, and can be adequately handled by fitting functions (see below). When $\eta-1\ll1$, we have a Dirac-delta distribution, and a better choice at all frequencies can be found using the low-frequency calculation (see Equation~\ref{eq:Omega_low_eta}). In other words, although the intermediate-frequency calculation is strictly valid only for $\eta\gg1$, we can adequately handle the error of this assumption using the freedom of the fitting functions below and the correct low-frequency limit derived above.

\subsubsection{High Frequencies \texorpdfstring{:$x_1\gg\eta^2$}{}}\label{sec:Psi}

In the high-frequency regime, we cannot solve for $J_{\rm pl}(p, x_1, x_\infty)$ exactly. Instead, we use the method of steepest descent (also referred to as Laplace's method) to approximate the integral. Taking the high-frequency limit of the synchrotron function $\tilde{F}(x \gg 1)$ (Appendix~\ref{appendix:synchrotron_function}), we have
\begin{equation}\label{eq:J_high_freq}
 J_{\rm pl}(p, x_1\gg1, x_\infty\gg1) \simeq \frac{\pi}{2}\int_{x_\infty}^{x_1} dx \,\,e^{-x}\, x^{(p-3)/2} \,\left(1-\sqrt{x_\infty/x}\right)^{p-2}.
 \end{equation}
We write the integral on the right-hand side as
\begin{equation}\label{eq:psi_p}
 \psi_p(q, x_\infty) \equiv \frac{\pi}{2}\int_{x_\infty}^{x_1} dx \,\,e^{-x}\, x^q \,\left(1-\sqrt{x_\infty/x}\right)^{p-2}.
 \end{equation}
 For $J_{\rm pl}$, we are interested only in $q=(p-3)/2$, but this generalized parameterization of $\psi_p$ will be useful below. The typical approach to the method of steepest descent / Laplace's method is derived by expanding an exponential factor in the integrand to second order about a maximum (see e.g., Chapter~6 of \citealt{BenderOrszag78}). This approach does not work well in the present case, since the integrand is strongly asymmetric and thus not approximated well as a Gaussian. Instead, we follow the procedure outlined in Problem~6.24 of  \cite{BenderOrszag78}. Changing variables to $t = \sqrt{x_\infty/x} $, the integral we are interested in can be written as
\begin{equation}\label{eq:psi_p_t}
 \psi_p(x_\infty) = \pi \,x_\infty^{(p-1)/2}\int_{1/\sqrt{\eta}}^{1} dt \,(1-t)^{p-2} \,\,e^{-x_\infty/t^2}\, t^{-p}.
 \end{equation}
Defining $\phi(t) = -1/t^2$, $g(t)= t^{-p}$, and  $\lambda = x_\infty$, this integral is of the general form
\begin{equation}\label{eq:psi_t}
 \int_{a}^b dt \,(b-t)^{\alpha} \,\,e^{\lambda\phi}\, g(t),
 \end{equation}
where $\alpha>-1$, $\lambda\gg1$, $g(b)=1$, and $\phi'(b) > 0$. The function $\phi(t)$ is assumed to take its maximum value for $a\leq t\leq b$ at $t=b$, though the derivative does not vanish at that point. The function $g(t)$ is assumed to vary slowly near $t=b$, so that a Taylor expansion in $g$ need not be taken. Expanding about $t=b$, the integral may be written approximately as
 \begin{equation}
e^{\lambda\phi(b)} \int_{a}^b dt \,(b-t)^{\alpha} \,\,e^{\lambda\phi'(b) (b-t)}\,.
 \end{equation}
The integrand is steeply peaked at $t=b$, so we may safely send $a\to 0$. We now change variables to $s = b-t$, so that $t$ ranging from $0$ to $b$ corresponds to $s$ ranging from $0$ to $b$. The integral is now steeply peaked at $s=0$, so that the contribution from $s>0$ is negligible and we can send the upper bound to $\infty$. Thus, we find
 \begin{equation}
 e^{\lambda\phi(b)} \int_{0}^\infty ds \,s^{\alpha} \,\,e^{\lambda\phi'(b) s} =  \frac{\Gamma(\alpha + 1)}{\{\lambda\phi'(b)\}^{\alpha + 1}}e^{\lambda\phi(b)}.
 \end{equation}
Inserting this into Equation~(\ref{eq:psi_t}),
\begin{equation}\label{eq:psi_p_final}
 \psi_p(q,x_\infty) \approx \,\frac{\pi\,\Gamma(p-1)}{2^{p-1}}x_\infty^{q -p+2} e^{-x_\infty}.
\end{equation}

\subsubsection{\texorpdfstring{$J_{\rm pl}$}{J pl} Fitting Function}

The final fitting function for $J_{\rm pl}$ is constructed in two steps. First, we join the intermediate- and high-frequency limits (Equations~\ref{eq:Jpl_inter}, \ref{eq:psi_p_final}) together using the fitting function
 \begin{equation}\label{eq:Psi_p}
    \Psi_p(x_\infty) = A_1(p)\,\delta^{\Psi}_1(x_\infty) + \psi_p\left(\frac{p-3}{2}, x_\infty\right) \,\delta^{\Psi}_2(x_\infty),
\end{equation}
where
\begin{equation}\label{eq:delta1_psi}
    \delta^{\Psi}_1(y) = \exp\left[-a_1y^2  - a_2 y^{2/3}\right] ,
\end{equation}
\begin{equation}\label{eq:delta2_psi}
    \delta^{\Psi}_2(y) = \left(1-\exp\{-a_4 y\}\right)^{a_3}
\end{equation}
are terms that turn ``on''/``off'' ($\delta^\Psi = 1$ or $0$) at high/low frequencies to ensure an appropriate transition between the intermediate- and high-frequency limits. To obtain an accurate fit for $\Psi_p$ for the range $2<p\leq5$, it is necessary in this case to fit the coefficients $a_i$ to 4th-order polynomials $a_i(p) = \sum_{j=0}^4 \aleph_{j} \,p^j$. The values of $\aleph_{j}$ are given in Table~\ref{table:b_vals}.

\begin{table*}
\centering
% \noindent\makebox[\textwidth]{
\begin{tabular}{l|c|c|c|c|c}
     Constant & $\aleph_0$ & $\aleph_1$ & $\aleph_2$ & $\aleph_3$ & $\aleph_4$ \\
     \hline
     $a_1$ & \num{0.502} & \num{-0.287} &   \num{0.057} &  \num{-0.004} & 0 \\
     $a_2$ & \num{1.115} & \num{0.441} &   \num{-0.006} &  \num{0} & 0 \\
     $a_3$ & \num{-3.27} & \num{3.17} &   \num{-0.718} &  \num{0.072} & \num{-0.0022} \\
     $a_4$ & \num{-0.221} & \num{0.721} &   \num{-0.352} &  \num{ 0.065} & \num{-0.004} \\
\end{tabular}%}
\caption{The values of the fitting constants used to define $\Psi_p$ (Equation~\ref{eq:Psi_p}), which interpolates between the intermediate-frequency behavior $A_1$ and high-frequency behavior $\psi_p$. The fitting constants $a_i$ (defined in Equations~\ref{eq:delta1_psi},\ref{eq:delta2_psi}) are polynomials in $p$ with coefficients $\aleph_j$, such that  $a_i(p) = \sum_{j=0}^4 \aleph_{j} \,p^j$.}
\label{table:b_vals}
\end{table*}

Next, we fuse $\Psi_p$ to the low-frequency limit of $J_{\rm pl}$ (Equation~\ref{eq:Omega}). For this purpose, we introduce a fitting function of the form
\begin{equation}\label{eq:J_pl_fitted}
    J_{\rm pl}(p, x_1, x_\infty) = \Omega_p(\eta, x_\infty) S_1(p,\eta,x_1) + \Psi_p(x_\infty) S_2(p,\eta,x_1),
\end{equation}
where $S_1$ and $S_2$ are sigmoids. For $\eta\gg1$, $S_1$ and $S_2$ converge to 0 at high and low frequencies, respectively. For $\eta-1\ll1$, $\Omega_p$ is the correct limit for all frequencies, so we require $S_2(\eta-1\ll1) =0$ independent of frequency. Varying $p$, we find an adequate fitting function using
\begin{equation}\label{eq:S_1}
    S_1(p,\eta,x_1) =  \exp\left[-\alpha_1 x_1^{\alpha_2} \exp\right(-\frac{\alpha_3}{\eta^2-1}\left)\right]
\end{equation}
\begin{equation}\label{eq:S_2}
    S_2(p,\eta,x_1) = \left[1-S_1(p,\eta,x_1)\right]^{\alpha_4} ,
\end{equation}
where the coefficients $\alpha_i$ are
\begin{equation}\label{eq:alpha_1}
    \alpha_1 =  -0.03\,p^3 + 0.45\,p^2 - 2.29\,p + 4.8,
\end{equation}
\begin{equation}\label{eq:alpha_2}
    \alpha_2 = 0.622   + 0.347 p^{2/3}  -0.017 p^{4/3},
\end{equation}
\begin{equation}\label{eq:alpha_4}
    \alpha_3 = 1 + (0.1p-0.71)e^{-(\eta^2-1.1)^2}.
\end{equation}
\begin{equation}\label{eq:alpha_3}
    \alpha_4 =   2 - 0.5e^{-(p-2)^2} + \left(0.538p - 1.23 + 0.5 e^{-(p-2)^2} \right) e^{-0.01(\eta^2-1.5)^2}
\end{equation}
The precise form of the fitting function has been chosen to minimize error for $2<p<5$ and $\eta\approx 1.1-1.2$, where neither $\Omega_p$ nor $\Psi_p$ captures the peak behavior precisely. We examine the fitting function for $J_{\rm pl}$ further in \S\ref{sec:PL_fitting_functions} after calculating similar fitting functions for the absorption coefficient.

\subsection{Absorption Coefficient: \texorpdfstring{$\alpha_{\nu, \rm pl}$}{alpha pl}}

The pitch-angle averaged absorption coefficient for a general electron distribution is \citep{RBL}
\begin{equation}\label{eq:alpha_nu}
    \alpha_{\nu} = \frac{1}{8\pi m_e\nu^2}\frac{\sqrt{3}e^3 B}{m_e c^2} \int_{\gamma_1}^{\gamma_2} d\gamma \,\, \frac{1}{\gamma^2}\left(\frac{\partial n}{\partial \gamma}\right) \frac{d}{d\gamma}\left[ \gamma^2 \tilde{F}(x) \right].
\end{equation}
Starting from Eq. 6.50 of \cite{RBL}, we have integrated by parts so that the derivative acts on the synchrotron function rather than on the distribution. The full integral used to compute the absorption coefficient runs from $\gamma = 1$ to $\infty$, or from zero to infinite kinetic energy. Assuming the distribution function is zero at these two points, then no boundary terms are picked up in the integration by parts. In the power-law case, the typical form for the absorption coefficient given in Equation 6.50 of \cite{RBL} and running from $\gamma=1$ to $\gamma=\infty$ leads to three terms: one from taking the derivative of $\gamma^{-p}$ and two from taking the derivatives of the discontinuous step functions $\Theta(\gamma-\gamma_1)$ and $\Theta(\gamma_2-\gamma)$ which are present in the full form of the power-law distribution. At high frequencies, the latter two terms can be ignored, but at low frequencies dropping these terms overestimates the absorption coefficient by a factor of $3(p+2)/4$ as compared to Equation 6.52 of \cite{RBL} (see footnote 6 of \citealt{GS02}). The form given in Equation~(\ref{eq:alpha_nu}), obtained by integrating by parts, does not suffer from these deficiencies, as the boundary terms vanish and no derivatives of step functions are taken. We reach a different conclusion than Appendix B of \cite{Warren22}, which argues that the \cite{GS02} form drops necessary boundary terms. As can be verified by a numerical calculation, this conclusion is incorrect: extra terms are needed only when analytically taking derivatives of the distribution, and not in the integrated-by-parts form of the absorption coefficient. The functional form used in Equation~(\ref{eq:alpha_nu}) matches onto the result that one would get from modeling the power-law distribution using a sharp but smooth cutoff at the endpoints (for example, by adding to the distribution a factor of $\exp[-(\gamma/\gamma_1)^{-n}] \exp[-(\gamma/\gamma_2)^{n}]$ for some $n \gg 1$) instead of a discontinuous jump via step functions.

Expanding the integrand using Equation~(\ref{eq:pl_dist_cooled}), setting $\gamma_2=\gamma_\infty$, and changing variables to $x$, this may alternatively be written
\begin{flalign}\label{eq:A_pl}
    \alpha_{\nu} & = -\frac{2\pi e}{3^{3/2}B}K_{\rm inj}\,\frac{n_e}{n_{e, \rm inj}} \, \mathscr{G}^{1-p} \left(\frac{\nu}{\nu_0}\right)^{-(p+4)/2}
    \int_{x_\infty}^{x_1} dx \,\, x^{(p+2)/2}\frac{d}{dx}\left(\frac{\tilde{F}(x)}{x}\right) \left(1-\sqrt{x_\infty/x}\right)^{p-2} & \nonumber \\
    & = \frac{2\pi e}{3^{3/2}B}K_{\rm inj}\,\frac{n_e}{n_{e, \rm inj}} \, \mathscr{G}^{1-p} \left(\frac{\nu}{\nu_0}\right)^{-(p+4)/2}
    A_{\rm pl}(p, x_1, x_\infty),
\end{flalign}
where 
\begin{equation}\label{eq:Apl}
    A_{\rm pl }(p,  x_1, x_\infty) \equiv 
    \int_{x_\infty}^{x_1} dx \,\, x^{(p+2)/2}\tilde{H}(x) \left(1-\sqrt{x_\infty/x}\right)^{p-2},
\end{equation}
using the function $\tilde{H}(x) = -\frac{d}{dx}(\tilde{F}(x)/x)$. Asymptotic limits and fitting functions for $\tilde{H}$ and its non-pitch-angle-averaged counterpart are provided in Appendix~\ref{appendix:synchrotron_function}. As before, we proceed by obtaining functions which are accurate at low and high frequencies. The second column of Table~\ref{table:Jpl_functions} provides a summary of the name of each limit of $A_{\rm pl}$ and where they can be located in the text.

\subsubsection{Low Frequencies \texorpdfstring{: $x_1\ll1$}{}}\label{sec:alpha_pl}

At low frequencies, we can solve for $A_{\rm pl}$ exactly. Following the steps given in \S~\ref{sec:omega_pl} with the replacement $x^{(p-3)/2}\tilde{F}(x) \to x^{(p+2)/2}\tilde{H}(x)$, we find the low-frequency limit 
\begin{equation}\label{eq:chi}
    \chi_p(\eta, x_\infty)  =  \frac{6}{3p+2} x_\infty^{(p+4)/2} \tilde{H}(x_\infty) \left\{\eta^{(3p+2)/3} \,_2F_1 (2-p, -2/3-p, 1/3-p, \eta^{-1}) - \frac{\Gamma(p-1)\Gamma(-p+1/3)}{\Gamma(-5/3)}\right\}.
\end{equation}
$\chi_p$ is valid at low frequencies for $\eta\gg1$ and at all frequencies for $\eta-1\ll1$. For $\eta\gg1$, we can set $x_\infty=0$ to obtain the limit 
\begin{equation}\label{eq:chi_low}
    \chi_p(\eta\gg 1, x_\infty\ll1)  
    \approx  \frac{4\tilde{F}_0}{3p+2} x_1^{(3p+2)/6}.
\end{equation}
For $\eta -1 \ll1$, the hypergeometric function may be expanded to give
\begin{equation}\label{eq:chi_low_eta}
     \chi_p (\eta-1\ll1, x_\infty\ll1)   
    \approx \,\frac{2}{p-1} x_1^{(p+4)/2} \tilde{H}(x_1) (\eta-1)^{p-1}.
\end{equation}

\subsubsection{Intermediate Frequencies \texorpdfstring{: $1 < x_1 < \eta^2$}{}}

We calculate the intermediate frequency limit by again setting $x_\infty=0$ everywhere, inserting the definition of $\tilde{H}(x)$, and explicitly calculating the pitch-angle averaging integration:
 \begin{flalign}\label{eq:B1}
    B_1(p)& = -\int_0^{\pi/2} d\alpha \,\,\sin^2{\alpha} \,\int_0^\infty dx \,\,\frac{d}{dx}\left(\frac{F(x/\sin{\alpha})}{x}\right) \,x^{(p+2)/2} & \nonumber \\
    & = \frac{p+2}{2}\int_0^{\pi/2} d\alpha \,\,\sin^2{\alpha} \,\int_0^\infty dx \,\,F(x/\sin{\alpha}) \,x^{(p-2)/2} & \nonumber \\
    & =\frac{p+2}{2}\int_0^{\pi/2} d\alpha \,\,\sin^{\frac{p+4}{2}}{\alpha} \,\int_0^\infty dx \,\,F(z) \,z^{(p-2)/2}& \nonumber \\
     & =\frac{\pi^{\frac{1}{2}}2^{\frac{p+2}{2}}}{p+4}\frac{\Gamma(\frac{p}{4} + \frac{3}{2})\Gamma(\frac{p}{4} + \frac{11}{6})\Gamma(\frac{p}{4} + \frac{1}{6})}{\Gamma(\frac{p}{4} + 1)}.
\end{flalign}
In the second line, we have integrated by parts, and in the third line changed variables to $z=x/\sin{\alpha}$. Any contribution from boundary terms at $x_1$ or $x_\infty$ is ignored here, an approximation valid for $\eta\gg1$. As for $J_{\rm pl}$, the intermediate-frequency regime is small for $\eta-1\sim 1$, and any error incurred from this assumption can be modified by the fitting function joining the low- and high-frequency limits. When $\eta-1\sim 1$, we reach a Dirac-delta distribution and every frequency is handled better by $\chi_p$ (see Equation~\ref{eq:chi_low_eta}). These limits are accounted for by the fitting functions below.

\subsubsection{High Frequencies \texorpdfstring{: $x_1\gg\eta^2$}{}}\label{sec:Apl_high_freq}

For high frequencies, we insert the high-$x$ limit for $\tilde{H}$ (Equation~\ref{eq:H_tilde_limits}) into the definition of $A_{\rm pl}$  to obtain\footnote{Note that the sub-leading term in the integrand proportional to $x^{(p-2)/2}$ does not give precisely the correct behavior, since we are ignoring an $\mathcal{O}(x^{(p-2)/2} )$  contribution from an asymptotic expansion of $\tilde{F}(x)$. Nevertheless, an accurate fitting function can be obtained using this approximate sub-leading term and the correct limit as $x\to\infty$ is preserved, so we include it in the fitting functions below.}
\begin{flalign}\label{eq:A_pl_high}
A_{\rm pl}(p, x_1,x_\infty) \simeq  \sigma_p(x_\infty) & \equiv \frac{\pi}{2}\int_{x_\infty}^{\infty} dx \,\,\left(x^{p/2} + x^{(p-2)/2}\right)\, e^{-x} \,\left(1-\sqrt{x_\infty/x}\right)^{p-2} & \nonumber \\ 
&= \psi_p
  \left(\frac{p}{2},x_\infty\right) + \psi_p
  \left(\frac{p-2}{2},x_\infty\right).
 \end{flalign}
For high frequencies, no new integral approximation is needed here: $\sigma_p$ may be written entirely in terms of $\psi_p(q,x_\infty)$ using Equation~(\ref{eq:psi_p}). Inserting the steepest descent approximation for $\psi_p$ (Equation~(\ref{eq:psi_p_final}), 
\begin{equation}\label{eq:chi_p_simplified}
\sigma_p(x_\infty) = \frac{2 \,\Gamma(p-1)}{2^{p-1}} x_\infty^{(p+2)/2} e^{-x_\infty} \left(1 + \frac{1}{x_\infty}\right).
 \end{equation}

\subsubsection{Final \texorpdfstring{$A_{\rm pl}$}{} Fitting Function}

With the intermediate-frequency constant $B_1$ and the high-frequency limit $\sigma_p$ in hand, we proceed as before and search for an analytic fitting function $\Sigma_p$ of the form
 \begin{equation}\label{eq:Sigma_p}
    \Sigma_p(x_\infty) = B_1  \,\delta^{\sigma}_1(x_\infty) +\sigma_p(x_\infty) \,\delta^{\sigma}_2(x_\infty),
\end{equation}
where
\begin{equation}\label{eq:delta1_sigma}
    \delta^{\sigma}_1(y) = \exp\left[-b_1y^2 -b_2 y^{2/3}\right]
\end{equation}
\begin{equation}\label{eq:delta2_sigma}
    \delta^{\sigma}_2(y) = \left(1-\exp[b_4 \,y^{1/3}]\right)^{b_3}.
\end{equation}
A good fit for $\Sigma_p$ can be found by fitting the constants to 4th-order polynomials,  $b_i(p) = \sum_{j=0}^4 \aleph_j \,p^j$. The fitted coefficients are listed in Table~\ref{table:e_vals}. 

\begin{table*}
\centering
\begin{tabular}{l|c|c|c|c|c}
     Constant & $\aleph_0$ & $\aleph_1$ & $\aleph_2$ & $\aleph_3$ & $\aleph_4$ \\
     \hline
     $b_1$ & \num{0.162}& \num{-0.124}&  \num{0.041} & \num{-0.0064} & \num{0.0004} \\
     $b_2$ & \num{-9.015}& \num{11.28}&  \num{-4.198} & \num{0.673} & \num{-0.036} \\
     $b_3$ & \num{43.43}& \num{ -50.35}&  \num{21.15} & \num{ -3.25} & \num{0.16} \\
     $b_4$ & \num{-2.566}& \num{6.366}&  \num{-2.51} & \num{ 0.428} & \num{-0.027} \\
\end{tabular}%}
\caption{Fitting constants used to define $\Sigma_p$ (Equation~\ref{eq:Sigma_p}). In order to capture the transition between intermediate and high frequencies, the constants (Equations~\ref{eq:delta1_sigma}, ~\ref{eq:delta2_sigma}) are fitted to polynomials in $p$, so that $b_i = \sum_{j=0}^{4} \aleph_j p^j$.}
\label{table:e_vals}
\end{table*}

The final fitting function for $A_{\rm pl}$ can be formed using $\chi_p$ and $\Sigma_p$ using
\begin{equation}\label{eq:A_pl_fitted}
    A_{\rm pl}(p, \eta, x_1) = \chi_p(\eta, x_\infty) S_3(p,\eta,x_1) + \Sigma_p(x_\infty) S_4(p,\eta,x_1).
\end{equation}
The sigmoids $S_3$ and $S_4$ satisfy the same respective limits as $S_1$ and $S_2$. A suitable fitting function for different values of $p$ and $\eta$ is
\begin{equation}\label{eq:S_1_A}
    S_3(p,\eta,x_1) =  \exp\left[-\beta_1 x_1^{\beta_2} \exp\right(-\frac{1}{(\eta^2-1)^{\beta_3}}\left)\right]
\end{equation}
\begin{equation}\label{eq:S_2_A}
    S_4(p,\eta,x_1) = \left[1-S_3(p,\eta,x_1)\right]^{\beta_4}.
\end{equation}
The fitting constants satisfy
\begin{equation}\label{eq:beta_1}
    \beta_1 =  0.077 + 29.16(p+10.71)^{-2} + \left[-0.063p + 0.253 - 29.16(p+10.71)^{-2}\right])e^{-10(\eta^2-1.1)^2},
\end{equation}
\begin{equation}\label{eq:beta_2}
    \beta_2 = 2 + 2.5 e^{-100(\eta^2-1.1)^2},
\end{equation}
\begin{equation}\label{eq:beta_3}
    \beta_3 =   1- 0.43e^{-500(\eta^2-1.01)^2},
\end{equation}
\begin{equation}\label{eq:beta_4}
    \beta_4 = 2.97p-3.13.
\end{equation}

\subsection{Comparison of Fitting Functions and Numerical Calculations}\label{sec:PL_fitting_functions}

\begin{figure*}
 \centering
  \includegraphics[width=0.9\textwidth]{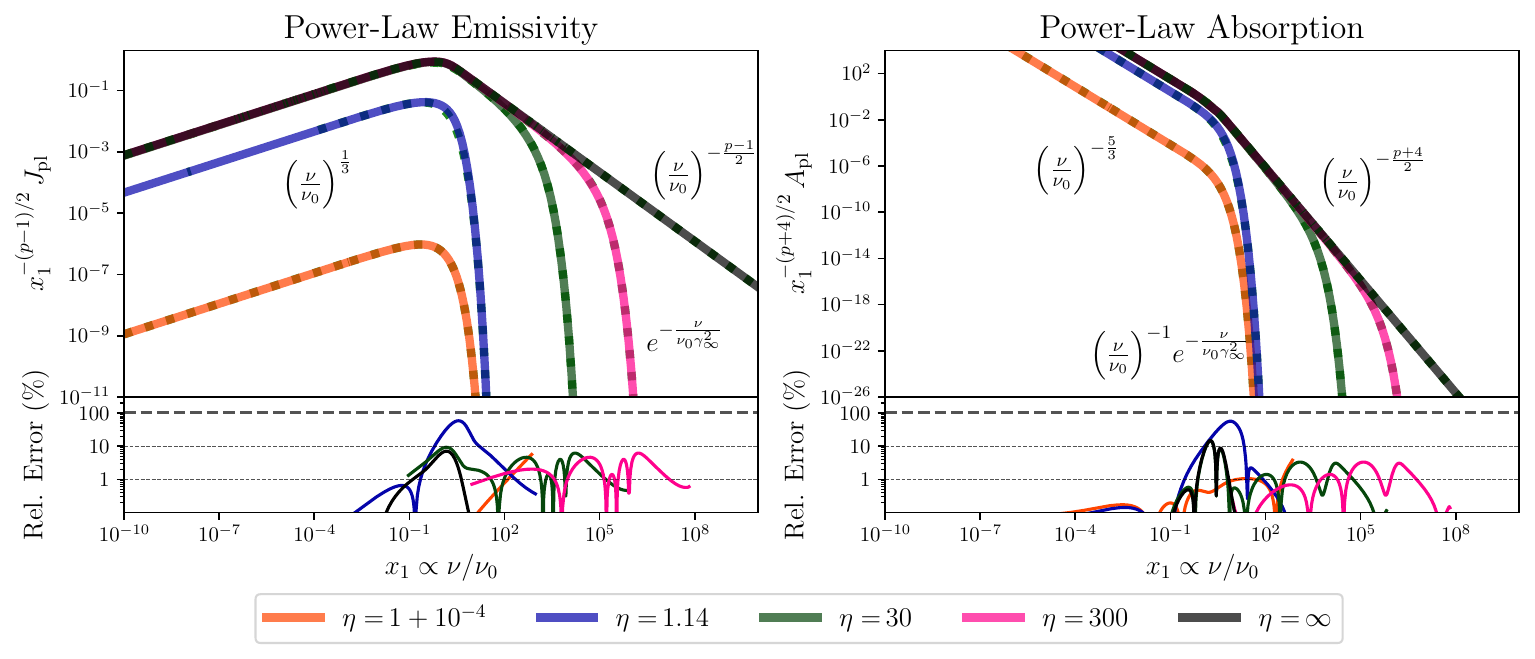}
  \caption{Non-dimensional forms of the power-law emissivity 
  $j_\nu \propto x_1^{-(p-1)/2} J_{\rm pl}(x_1)$ (Equation~\ref{eq:J_pl})
  and absorption coefficient 
  $\alpha_\nu \propto x_1^{-(p+4)/2} A_{\rm pl}(x_1)$ (Equation~\ref{eq:A_pl}) 
  compared to our fitting functions (squares and colored lines, respectively). 
  The emissivity and absorption coefficients are shown as a function of frequency $\nu\propto x_1 \equiv x(\gamma_1)$ (Equation~\ref{eq:x}) and for different values of $\eta \equiv \gamma_\infty/\gamma_1$, the ratio between the minimum and maximum electron Lorentz factors of the cooled distribution (Equation~\ref{eq:eta}). Approximate frequency scalings in each regime are shown in the top two panels. Note that the $\eta=\infty$ curve is numerically set at  $\eta=10^{10}$; at these frequencies the coefficients corresponding to $\eta=10^{10}$ and $\eta=\infty$ are indistinguishable. The bottom panels show the absolute value of the relative error between the fitting functions and the numerical calculations. The relative errors are maximized around $x_1 \sim 1$, where the low- and high-frequency parts of the fitting function cross over. The errors are highest ($58.2\%$ and $56.4\%$) for $\eta=1.14$; this value of $\eta$ was chosen to maximize the error of the fitting functions. For $\eta= 1 + 10^{-4}$, the maximum errors (occurring at high frequencies, where the overall coefficients are exponentially suppressed) are $5.6\%$ and $3.8\%$. For other values of $\eta$, the maximum errors are $9.3\%$ for $J_{\rm pl}$ and $14.7\%$ for $A_{\rm pl}$. Typical relative errors are at the percent level in each case.}
  \label{fig:PL_coeff}
  \end{figure*}

We can now test the ability of the fitting functions to accurately reproduce $J_{\rm pl}$ and $A_{\rm pl}$. In Figure~\ref{fig:PL_coeff} we first look at the emission and absorption coefficients with fixed $p=2.5$ and $\eta = 1+10^{-4}$, $1.14$, $30$, $300$, and $\infty$. As the emitting electrons cool, the value of $\eta$ decreases, and we may think of the different curves as representing the state of the radiation coefficients at different times. 

\begin{figure*}
 \centering
  \includegraphics[width=0.9\textwidth]{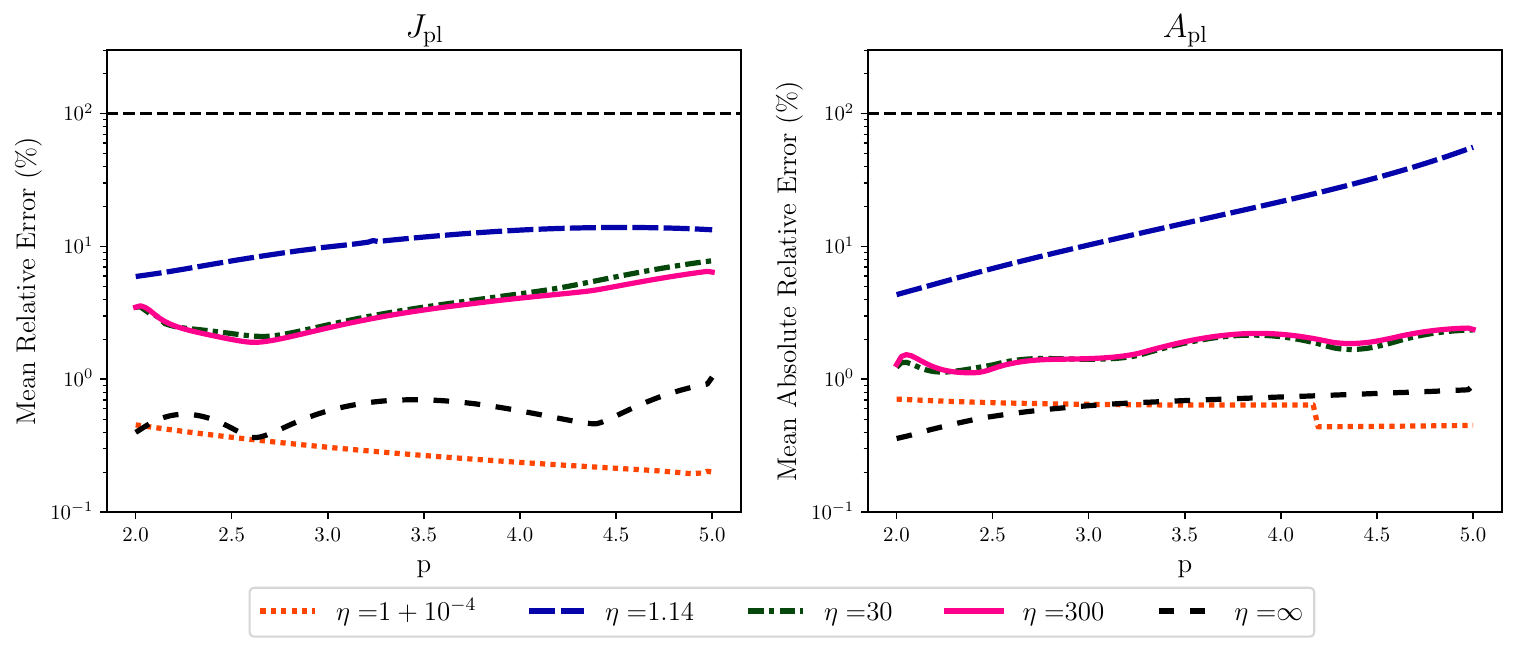}
  \caption{The mean of the absolute value of the relative errors in the fitting functions for $J_{\rm pl}(p, x_1,\eta)$ and $A_{\rm pl}(p, x_1,\eta)$ as a function of $p$ (within the range $2<p\leq5$ explored in this work) for five different values of $\eta$. To evaluate each curve on the same footing, the percent error is calculated using the formula $\frac{100}{n} \sum_i^n\,|1-f^i_{\rm pl, fitted}/f^i_{\rm pl, numerical}|$, where $n \leq 1000$ is the number of $x_1$ values between $10^{-10}$ and $10^{10}$ for which each function does not evaluate to zero. The fitting functions are best for $p\lesssim3.5$, which is the regime of greater relevance to typical astrophysical settings. The maximum errors in each panel are $13.9\%$ and $55.9\%$, respectively. In both cases, these values occur at $\eta=1.14$. The maximum error for other values of $\eta$ is at or below the percent level, indicating that on average, the fitting functions behave sufficiently well to act as a substitute for the full numerical integration.}
  \label{fig:PL_error}
  \end{figure*}

As $\eta$ falls, the intermediate regime of size $\eta$ becomes smaller and smaller, and the emissivity eventually resembles a rescaled form of the synchrotron function $\tilde{F}(x)$. For $\eta\gg1$ and $\eta-1\ll1$, the absolute value of the relative error in the bottom panels is under control, vanishing as $x_1\to0$ and $x_1\to\infty$. The maximum relative error peaks at the ten percent level near $x_1=1$, the transition between the low- and intermediate-frequency regimes. The relative error is highest for $\eta\approx1.1-1.2$, where neither the low- or high-frequency functions are especially accurate. This effect is higher for larger values of $p$ and more pronounced for the absorption coefficient, where the maximum relative error can rise above $100\%$. Fixing a value of $p$, the mean relative error for all values of $x_1$ is nevertheless typically below $10\%$ (Figure~\ref{fig:PL_error}). 

For $\eta=1.14$, the error in Figure~\ref{fig:PL_error} is highest, reaching a maximum at $55.9\%$. The value of $1.14$ was chosen since it maximizes the mean relative error; other values of $\eta$ have smaller errors. The instances of the fitting function with larger mean errors are clustered narrowly around $\eta\simeq 1.05$ and occur predominantly for $p>3.5$. For more typical values of $p$, the fitting functions behave reasonably well in all cases.

\subsection{Comparison to Chopped-Off Distribution}\label{sec:FO14}

To conclude our discussion of cooled synchrotron emission from power-law electrons, we compare our results to those obtained from a ``chopped-off" power-law between Lorentz factors $\gamma_1$ and $\gamma_2 = \gamma_\infty$. For $\eta\gg1$, the precise cooled distribution function (Equation~\ref{eq:pl_dist_cooled}) is qualitatively similar to a chopped power-law (see Figure~\ref{fig:distributions}), but the number of high-energy electrons is truncated smoothly by factors of  $1-\gamma/\gamma_\infty$ instead of cut off sharply at $\gamma_2$. Here we examine in detail the error induced by ignoring the smooth-cutoff factors in favor of the simpler sharp-cutoff model often used to model cooling in one-zone models (e.g., \citealt{FO14}). This comparison allows us to examine the significance of the cooling-induced changes to the functional form of the distribution, as opposed to the distinct ---but related---effect of cooling which simply decreases the maximal Lorentz factor. 

\begin{figure*}
 \centering
  \includegraphics[width=0.9\textwidth]{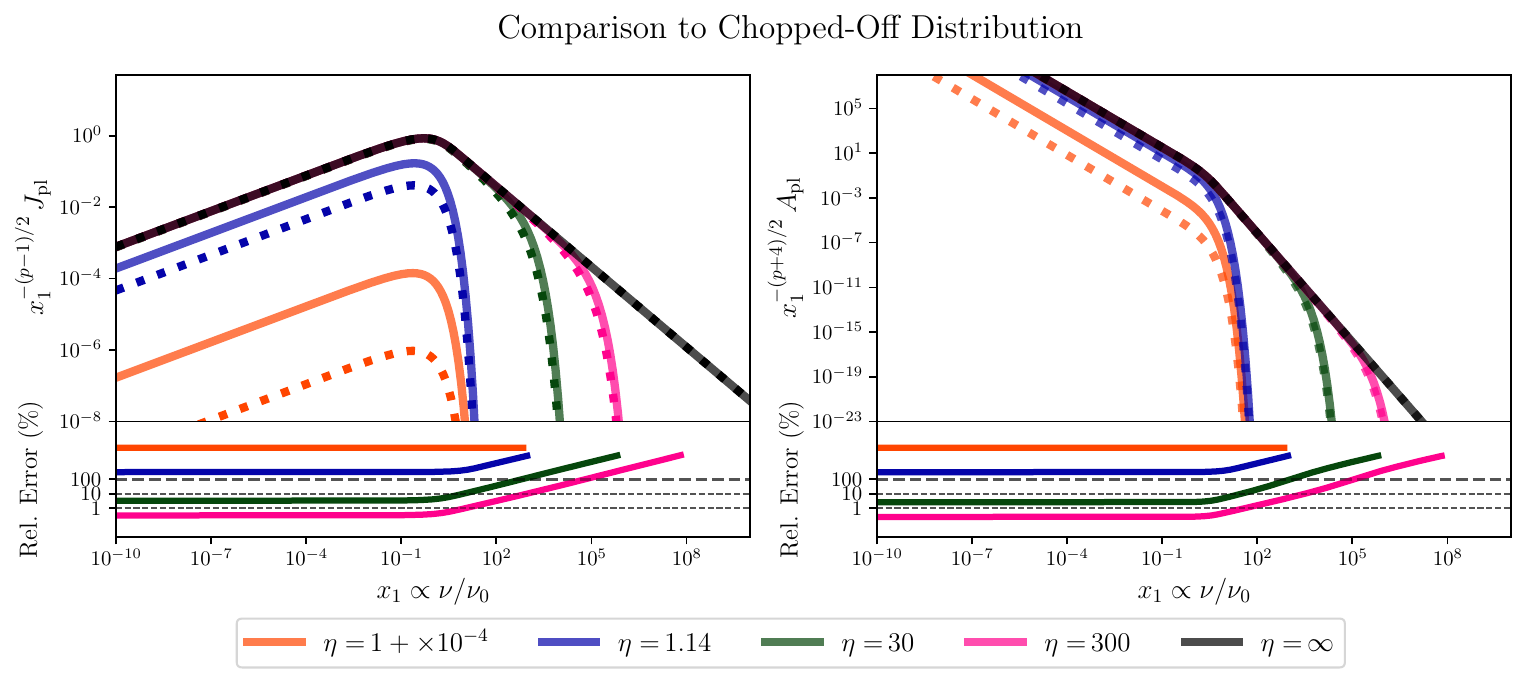}
  \caption{Power-law emission and absorption coefficients from the smoothly cutoff cooled distribution treated in this section compared to the result of using a chopped-off distribution; that is, a power-law abruptly truncated at Lorentz factor $\gamma_\infty$ as opposed to the smoothly cut off distribution found by solving the electron cooling ODE (Equation~\ref{eq:pl_dist_cooled}). For large values of $\eta = \gamma_\infty/\gamma_1$, the chopped-off distribution is a good approximation for the radiation coefficients at low frequencies. At high frequencies where cooling is significant ($x_2\gtrsim 1$), the predictions of the chopped-off distribution do not accurately match calculations that use exact forms for the cooled electron distribution functions (note, however, that the discrepancies are largest on the exponential tail where emission/absorption are suppressed). As $\eta$ becomes smaller, the discrepancy becomes greater at lower frequencies, and for $\eta\to 1$ the relative error becomes increasingly large.
  }
  \label{fig:FO14_comparison}
  \end{figure*}

In Figure~\ref{fig:FO14_comparison}, we plot five curves comparing the sharp-cutoff formalism, obtained by computing Equations~(\ref{eq:J_pl}, \ref{eq:A_pl}) numerically after dropping factors of $1-\sqrt{x_\infty/x}$, to the smooth-cutoff formalism, which is calculated by a full numerical integration of Equations~(\ref{eq:J_pl}, \ref{eq:A_pl}). 
The impact of using the cooled smooth-cutoff distributions over the sharp-cutoff distributions depends on frequency. For large values of $\eta$ and high frequencies, the difference in the treatment of the high-energy tail causes a corresponding difference in the  exponential tail of the radiation coefficients. For sufficiently high $x_1$, the smooth- and sharp-cutoff radiation coefficients never match, as the smooth-cutoff coefficients falls to zero faster than implied by the sharp-cutoff distribution. The steepest descent approach used in this work captures the high-frequency behavior of the emission and absorption coefficients much better, as can be seen in the analogous case in Figure~\ref{fig:PL_coeff}. For small frequencies, both distributions recover the same results, which is expected since for large $\eta$ cooling mainly affects the highest-energy electrons.

As $\eta\to 1$, the smooth-cutoff formalism applied to the local cooling case become very inaccurate at all frequencies. In this limit, $\gamma_1$ and $\gamma_2$ are comparable, and the inclusion of the term $ \propto (1-\gamma/\gamma_\infty)^{p-2}$ is highly relevant at all frequencies. More technically, the sharp-cutoff distribution assumes that Equation~(\ref{eq:Omega_high_eta}) applies at low frequencies for all values of $\eta$, whereas the full local treatment makes clear that for small $\eta$ the correct emissivity is given by Equation~(\ref{eq:Omega_low_eta}).
For applications to local cooling, the sharp-cutoff distribution is only useful at low frequencies and when $\gamma_\infty\gg\gamma_1$. We conclude that the smooth-cutoff fitting functions are, in general, needed to correctly handle local power-law cooling.

\section{Fitting Functions: Thermal Distribution}\label{sec:th_dist}

We now consider a relativistic thermal distribution of electrons which initially extends from Lorentz factors $1$ to $\gamma_3=\gamma_\infty$. As time increases, adiabatic cooling lowers the effective temperature of the injected distribution. As the highest-energy electrons cool radiatively, $\gamma_3$ also falls in time and we are left with a cooled thermal distribution. 
We note that electrons are assumed to be thermalized only at the time of injection, and that during the subsequent time evolution (as electrons advect downstream of the shock) there is no re-heating or re-thermalization of these electrons. In particular, collisional heating is assumed to be inefficient at times $t>t_{\rm inj}$. 
This leads to the `cooled thermal distribution' derived in \S~\ref{sec:distributions}, which technically is not an equilibrium thermal (Maxwellian) distribution.
We further discuss this issue in \S~\ref{sec:conclusion}. 
In this section, we consider the emissivity and absorption resulting from this distribution, providing analytic fitting functions for each case.

Inserting the cooled thermal distribution (Equation~\ref{eq:therm_dist_cooled}) into the definition of the emission coefficient, we find, assuming $\gamma_3=\gamma_\infty$,
\begin{flalign}
    j_{\nu, \rm th} & = \frac{1 }{4\pi } \int_{1}^{\gamma_3} d\gamma \left(\frac{\partial n}{\partial \gamma}\right)_{\rm th} \tilde{P}_e & \nonumber \\
    &\simeq \frac{\sqrt{3}e^3 B }{4\pi m_e c^2} \frac{L_{\rm inj}}{2} \frac{n_e}{n_{e, \rm inj}} \int_{1}^{\gamma_\infty} d\gamma \,\,
     \frac{\gamma^2}{\bar{\Theta}^3}\, \frac{\tilde{F}(x)}{(1- \gamma/\gamma_{\infty})^4}\,\exp \left[ -\frac{1}{\bar{\Theta}} \frac{\gamma}{1- \gamma/\gamma_{\infty}}\right] & \nonumber \\
    & \equiv \frac{\sqrt{3}e^3 B }{4\pi m_e c^2} \frac{L_{\rm inj}}{2\bar\Theta} \frac{n_e}{n_{e, \rm inj}} J_{\rm th}(y, \etath ),
\end{flalign}
where we have defined the effective local temperature
$\bar{\Theta} = \Theta/\mathscr{G}$ and, since we are primarily interested in the emission from ultra-relativistic electrons with Lorentz factors $\gamma \gg 1$, we have ignored the square-root term equal to the electron velocity. Changing variables to $z=\gamma/\bar{\Theta}$, the function $J_{\rm th}$ for which we will derive a fitting function may be written
\begin{flalign}\label{eq:Jth}
    J_{\rm th}(y, \etath)
     & \simeq  \int_{0}^{\etath} dz \,\,
     z^2\, \frac{\tilde{F}(y/z^2)}{(1- z/\etath)^4}\,\exp \left[ - \frac{z}{1- z/\etath}\right], &\notag\\
      & \simeq  \int_{0}^{\infty} ds \,\,
     s^2\, e^{-s} \,\tilde{F}\left(\frac{y\,(1 + s/\etath)^2}{s^2}\right),
\end{flalign}
where $y = (\nu/\nu_0)/\bar{\Theta}^2 \equiv \nu/\
\nu_{\rm \bar\Theta}$ 
is a dimensionless frequency coordinate
and $\etath = \gamma_\infty/\bar{\Theta}$, the maximum allowed value of $z$, governs the importance of cooling. In the second line, we make a change of variables to $s = z/(1-z/\etath)$. Both forms of $J_{\rm th}$ will be used to derive asymptotic limits below. The formulation in terms of $s$ is more useful for numerical integration. We eliminate any explicit dependence on $\bar{\Theta}$ by assuming $\bar\Theta\gg1$.\footnote{Without this assumption, the integral defining $J_{\rm th}$  (and $A_{\rm th}$ below) runs from  $1/\bar{\Theta}$ to $\etath$.} 
This assumption is justified even for $\bar{\Theta}\sim 1$, because electrons with $z=1/\bar{\Theta}$ (in other words, at $\gamma=1$) contribute negligibly to the emission.

The cooled thermal absorption coefficient is analogously
\begin{flalign}
    \alpha_{\nu, \rm th} & =  \frac{1}{8\pi m_e\nu^2}\frac{\sqrt{3}e^3 B}{m_e c^2}\int_{1}^{\gamma_3} d\gamma \,\, \frac{1}{\gamma^2}\left(\frac{\partial n}{\partial \gamma}\right)_{\rm th} \frac{d}{d\gamma}\left(\gamma^2 \tilde{F}(y/z^2)\right) & \nonumber \\
    & \equiv \frac{\pi e}{3^{3/2} B} L_{\rm inj} \,\frac{n_e}{n_{e, \rm inj}} \frac{2}{\bar{\Theta}^5} A_{\rm th}(y, \etath),
\end{flalign}
with the implicit definition

\begin{flalign}\label{eq:Ath}
    A_{\rm th}(y, \etath)
      & \simeq  \int_{0}^{\etath} dz \,\,
     z^{-3}\, \frac{\tilde{H}(y/z^2)}{(1- z/\etath)^4}\,\exp \left[ - \frac{z}{1- z/\etath}\right]&\\
     &        \simeq  \int_{0}^{\infty} ds \,\,
     s^{-3}\, (1 + s/\etath)^5\,e^{-s} \,\tilde{H}\left(\frac{y\,(1 + s/\etath)^2}{s^2}\right).
\end{flalign}
For cooled thermal distributions, we only need to worry about two frequency regimes, $y \ll  \ymax$ and $y\gg \ymax$, separated by a transition frequency $\ymax$. In each regime, we make use of the asymptotic limits of the relevant synchrotron functions (Appendix~\ref{appendix:synchrotron_function}), with the final fitting function interpolating between these extremes. The transition frequency $y_t = z_{\rm pk}^2$ may be defined in terms of the peak value $z_{\rm pk}$ of the distribution function, with the form chosen so that the argument of the synchrotron function is unity when $y=y_t$ and $z=z_{\rm pk}$. For $\etath\gg1$, $\gamma_{\rm pk} \sim \bar{\Theta}$ and therefore $z_{\rm pk} \sim 1$. By contrast, as we will see below, for small $\etath$, the distribution peaks at $z \simeq \etath$, and thus  $\ymax \sim \etath^2$. We choose $y_t$ such that $y_t \sim \min ( 1, z_\infty^2 )$ obeys these two limits, and make the specific choice
\begin{equation}\label{eq:y_t}
    \ymax(\etath) =  
    -{\ln \left( e^{-100} + e^{-100\,\etath^2} \right)}/{100} ,
\end{equation}
which acts as a smooth minimum function.

\subsection{Low Frequencies \texorpdfstring{: $y\ll \ymax$}{}}\label{sec:pi}

For low frequencies, we use $\tilde{F}(x\ll1) = \tilde{F}_0 x^{1/3}$ and $\tilde{H}(x\ll1) = \frac{2\tilde{F}_0}{3} x^{-5/3}$ (see Appendix~\ref{appendix:synchrotron_function}). In this regime, it is convenient to use the variable $s$, which yields
\begin{equation}
    J_{\rm th}(y\ll \ymax)
     \simeq  \tilde{F}_0 y^{1/3}\int_{0}^{\infty} ds \,\,
     s^{4/3}\, (1+s/\etath)^{2/3} \,e^{-s}
\end{equation}
\begin{equation}
     A_{\rm th}(y\ll \ymax)
     \simeq  \frac{2\tilde{F}_0}{3} y^{-5/3}\int_{0}^{\infty} ds \,\,
     s^{1/3}\, (1+s/\etath)^{5/3} \,e^{-s}.
\end{equation}
These integrals can be expressed in terms of the confluent hypergeometric function of the second kind $U$,\footnote{For $\etath\ll1$, the confluent hypergeometric functions used in the thermal coefficients satisfy the limits $U(7/3, 4, \etath\ll1) \simeq -27/4\,\Gamma(-2/3)\etath^3 -9/4\,\Gamma(-2/3)\etath^2$ and $U(4/3, 4, \etath\ll1) \simeq 27/5\,\Gamma(-5/3)\etath^3 +9/2\,\Gamma(-5/3)\etath^2$. For $\etath\lesssim10^{-3}$, it can be necessary to use these alternative limits directly to ensure numerical stability.}
\begin{equation}\label{}
    J_{\rm th}(y\ll \ymax)
     \simeq    \tilde{F}_0\, y^{1/3}\, \etath^{7/3} \,\Gamma(7/3)\,\, U(7/3, 4, \etath) 
\end{equation}
\begin{equation}\label{}
     A_{\rm th}(y\ll \ymax)
     \simeq  \frac{2\tilde{F}_0}{3}\, y^{-5/3}\, \etath^{4/3} \,\Gamma(4/3)\,\, U(4/3, 4, \etath) .
\end{equation}
Similarly to the low-$\eta$ regime for power-law electrons, when $\etath\ll1$ the cooled thermal distribution starts to behave like a Dirac delta in energy (see Figure~\ref{fig:distributions}). Thus, the integrals giving the emission and absorption coefficients are (respectively) proportional to $\tilde{F}(y/\etath^2)$ and $\tilde{H}(y/\etath^2)$. Making use of Equations~(\ref{eq:F_tilde}, \ref{eq:H_tilde}) and using the $y\ll1$ limit derived above, the emission and absorption coefficients for $\etath\ll1$ are simply 
\begin{equation}\label{eq:pi_J}
    J_{\rm th}(\etath\ll1)
     \simeq    \tilde{F}(y/\etath^2) \, \etath^{3} \,\Gamma(7/3)\,\, U(7/3, 4, \etath) \equiv \Pi_J(y,\etath)
\end{equation}
\begin{equation}\label{eq:pi_A}
     A_{\rm th}(\etath\ll1)
     \simeq  \tilde{H}(y/\etath^2)\,\, \etath^{-2} \,\Gamma(4/3)\,\, U(4/3, 4, \etath) \equiv \Pi_A(y,\etath).
\end{equation}
These forms of $J_{\rm th}$ and $A_{\rm th}$ are valid when $y\ll1$ or for all frequencies when $\etath\ll1$.

\subsection{High Frequencies \texorpdfstring{: $y\gg \ymax$}{}}\label{sec:upsilon}

For high frequencies, we again use the method of steepest descent. For $\etath=\infty$, cooling is unimportant and the emission and absorption coefficients reduce to the form calculated by \cite{Mahadevan96}. In this limit, $J_{\rm th} \to y I'(y)$ and $A_{\rm th}\to I'(y)/2y$, where $I'(y) = 4.0506 y^{-1/6} (1 + 0.4 y^{-1/4} + 0.5316 y^{-1/2}) \exp (-1.8899 y^{1/3})$ is a fitting function given in Equation 32 of \cite{Mahadevan96}, accurate to within $2.7\%$. The limit of $I'(y)$ for $y\gg1$ is derived using the method of steepest descent, for which the saddle point occurs at $z_s = (2y)^{1/3}$ \citep{Petrosian81}. To use the method of steepest descent for finite $\etath$, we must consider alterations both to the overall integrals and to the location of the saddle point $z_s$. Using the asymptotic expansions of the synchrotron functions for large $x$, we find $J_{\rm th}(y, \etath) = \frac{\pi}{2}\xi_2(y, \etath)$ and $A_{\rm th}(y, \etath) = \frac{\pi}{2y}\xi_{-1}(y, \etath)$, where the same general function $\xi_q$ may be used in both cases,
\begin{equation}\label{eq:xi}
    \xi_q(y, \etath) \equiv \int_0^{\etath} dz \,\, z^q \exp\left[-\frac{z}{1-z/\etath} - \frac{y}{z^2} -4 \ln \left( 1-\frac{z}{\etath} \right)\right].
\end{equation}
This integral  may be estimated to sufficient accuracy using the typical steepest descent formula:
\begin{equation}\label{eq:xi_sd}
    \xi_q(y, \etath) \sim \sqrt{\frac{2\pi}{-\phi''(y,\etath,z_s)}} z_s^q \,e^{\phi(y,\etath,z_s)},
\end{equation}
where
\begin{equation}\label{eq:phi}
    \phi(y,\etath, z) = -\frac{z}{1-z/\etath} - \frac{y}{z^2} -4 \ln \left( 1-\frac{z}{\etath} \right),
\end{equation}
\begin{equation}\label{eq:phi_double_prime}
    \phi''(y,\etath, z) = -\frac{2}{\etath(1-z/\etath)^3} - \frac{6y}{z^4}  +\frac{4}{(z-\etath)^2},
\end{equation}
and $z_s$ is the local minimum at which $\phi'(z_s)=0$. The condition that the derivative of $\phi$ equal 0 leads to a quartic equation for $z_s$,
\begin{equation}\label{eq:quartic}
   \frac{4}{\etath^2} z_s^4 + \left(1-\frac{4}{\etath}\right) z_s^3 - \frac{2y}{\etath^2}z_s^2 + \frac{4y}{\etath} z_s -2y =0.
\end{equation}
This equation may be solved numerically if desired. For the purposes of deriving fully analytic fitting functions, instead of an exact solution, we implement a fitting function for the positive real root $z_s$. It is useful at this point to define the characteristic frequency
\begin{equation}
    y_0 = \etath^3/2.
\end{equation}
For $y\ll y_0$, all terms are nearly zero except for the first cubic term and the final constant, leading to $z_s\sim (2y)^{1/3}$, the result given by \cite{Petrosian81}. When $y\gg y_0$, we have $z_s \simeq \etath$. Expanding Equation~(\ref{eq:quartic}) in terms of the small number $\epsilon =1-z_s/\etath$, we can obtain a quadratic equation with the positive root\footnote{Note that one of the terms in the expansion, $-3\epsilon^2$, has been dropped. Since $y\gg y_0$, this term is typically unimportant; ignoring it improves the numerical convergence of the approximate solution when $y\sim y_0$.}
\begin{equation}\label{z_s_high}
    \epsilon(y\gg  y_0) = \frac{-\frac{4}{\etath} -3 +\sqrt{(\frac{4}{\etath}+3)^2 +4(\frac{y}{ y_0} - \frac{12}{\etath})}} {2(\frac{y}{ y_0} - \frac{12}{\etath})}.
\end{equation}
To leading order for $y\gg y_0$,  $\epsilon \sim (y/y_0)^{-1/2}$. For $\etath\ll1$, only the latter solution with $z_s = \etath(1-\epsilon)$ is relevant. 
Overall, we find that a suitable approximation for the saddle-point solution to Equation~(\ref{eq:quartic}) $z_s$ for arbitrary $y$ and $\etath$ is
\begin{equation}\label{eq:zs_fit}
    z_{s, \rm fitted} = \left[\left(\etath-\etath\epsilon\right)^{-2.019} +\left(\frac{2.7\times10^4}{ y_0^4} + (2y)^{1/3}\right)^{-2.019 } \right]^{-0.495}.
\end{equation}
Using this form for $z_s$ in Equation~(\ref{eq:xi_sd}) completes the construction of an approximation for $\xi_q$.

\subsection{Final Fitting Functions}

\begin{figure*}
 \centering
  \includegraphics[width=0.9\textwidth]{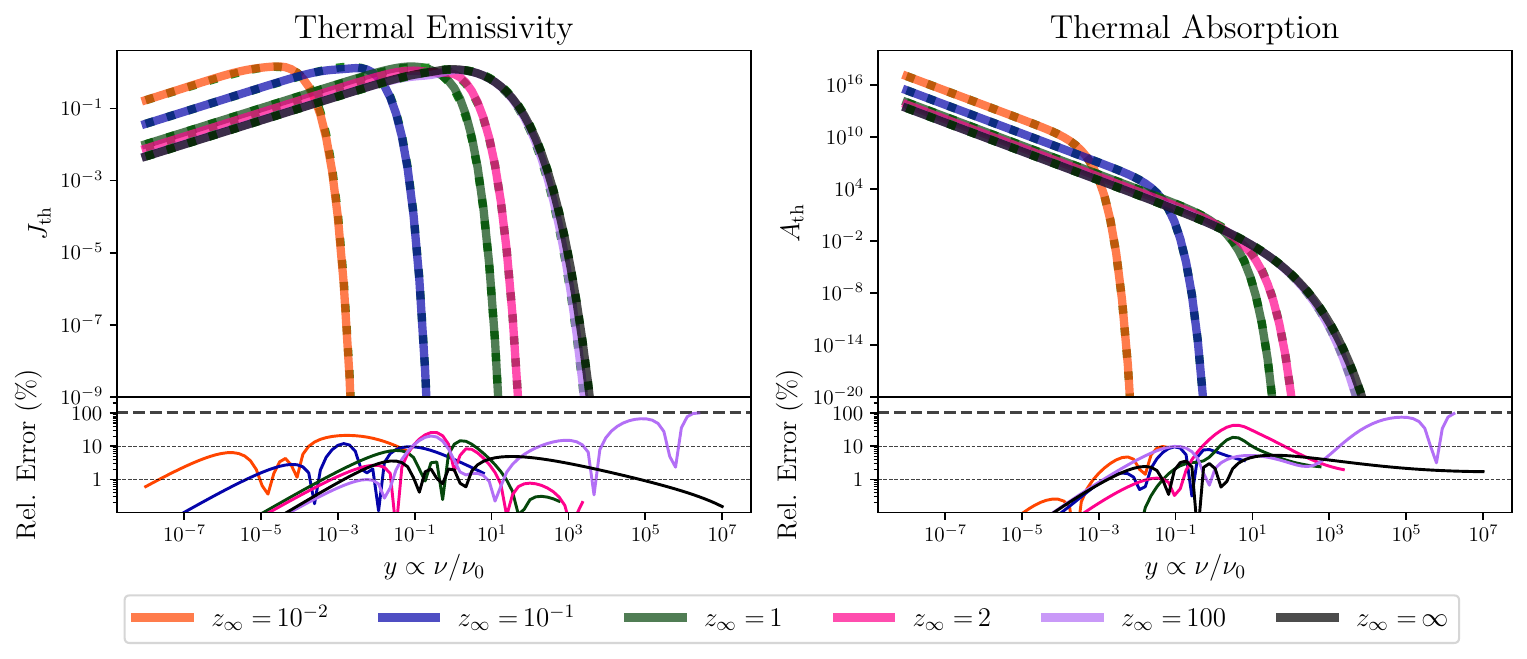}
  \caption{Cooled thermal emission and absorption coefficients (Equations~\ref{eq:Jth},\ref{eq:Ath}) for different values of $\etath$ and as a function of $y\propto\nu/\nu_0$. In the top panel, the values of the radiation coefficients are calculated in two ways, corresponding to the exact numerical calculation (squares) and approximate fitting functions (solid lines). The curve labeled $\etath=\infty$ is evaluated at  $\etath=10^8$. As for the power-law case, $\etath=10^8$ is a good proxy for $\eta=\infty$: at these frequencies the difference in the radiation coefficients between the two curves is unimportant. The bottom panel shows the relative error between the two calculations. Because the fitting functions use an approximate form for the saddle point $z_s$ (Equation~\ref{eq:zs_fit}), the error climbs at large frequencies when the coefficients are exponentially suppressed. Because the coefficients are strongly suppressed at these frequencies, this error is inconsequential. At $y\sim y_{\rm t}$ (Equation~\ref{eq:y_t}), near the turnover between low- and high-frequency behavior, the relative error is always maximum at the $10$--$20\%$ level, with typical values $\sim1\%$.}
  \label{fig:TH_coeff}
  \end{figure*}

\begin{figure*}
 \centering
  \includegraphics[width=0.5\textwidth]{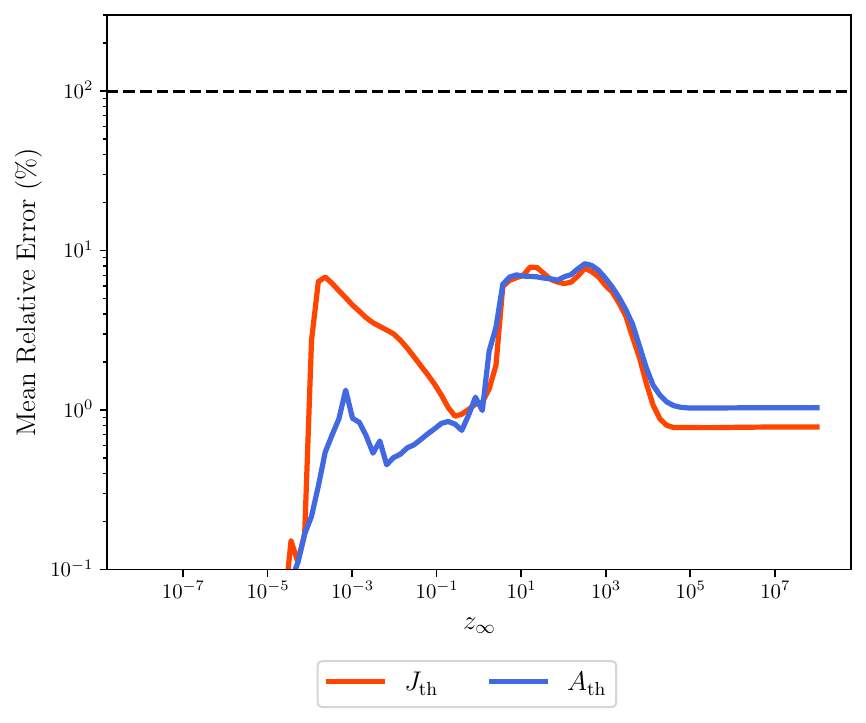}
  \caption{Mean of the absolute value of the relative errors between exact numerical forms approximate fitting functions for the thermal radiation coefficients $J_th(y, \etath)$ and $A_th(y, \etath)$. For each $\etath\in(10^{-8},10^{8})$, the percent mean relative error is calculated as $\frac{100}{n} \sum_i^n\,|1-f^i_{\rm pl, fitted}/f^i_{\rm pl, numerical}|$, where $n \leq 400$ is the number of $y$ values between $10^{-20}$ and $10^{8}$. The maximum errors for each curve are $7.9\%$ ($J_{\rm th}$) and $8.3\%$ ($A_{\rm th}$). The mean relative error peaks between $\etath\sim10^{-4}-10^4$, below and above which the error remains at the percent level or below.}
  \label{fig:TH_error}
  \end{figure*}

Using the low-frequency and high-frequency limits derived in the last two subsections, we can find fitting functions for the thermal emission and absorption coefficients. We choose the functional forms
\begin{equation}\label{eq:J_th_fitted}
    J_{\rm th}(y,\etath) = \Pi_J(y,\etath) e^{-\lambda_1(\etath)\,(\frac{y}{y_t})^{1.015} \,\,\zeta(\etath)} + \left[ 1 + \rho(\etath)\, \sqrt{y_t/y}\right] \frac{\pi}{2}\, \xi_2(y,\etath) \left[ 1- e^{-\lambda_2(\etath)\,(\frac{y}{y_t})^{1.015} \,\,\zeta(\etath)} \right],
\end{equation}
\begin{equation}\label{eq:A_th_fitted}
    A_{\rm th}(y,\etath) = \Pi_A(y,\etath) e^{-\mu_1(\etath)\,(\frac{y}{y_t})^{1.015} \,\,\zeta(\etath)} + \left[ 1 + \rho(\etath)\, \sqrt{y_t/y}\right] \frac{\pi}{2}\, \xi_{-1}(y,\etath) \left[ 1- e^{-\mu_2(\etath)\,(\frac{y}{y_t})^{1.015} \,\zeta(\etath)} \right].
\end{equation}
Suitable values for the constants may be found by numerical experimentation; here we find a good fit by allowing the constants to be functions of $\etath$. Defining the auxiliary variable $\kappa \sim y_t(z_\infty)/\mathrm{\rm min}(\frac{1}{2}z_\infty^3,1)$ such that\footnote{This amounts to using a slightly altered ``effective" transition frequency in these cases, which improves the fitting procedure for the pitch-angle averaged coefficients. In contrast, the perpendicular pitch-angle functions presented in Appendix~\ref{app:perp_pitch_angle} make use of the correct $y_t$.}
\begin{equation}
    \kappa(z_\infty) = -100 \,y_t(z_\infty) / \ln \left( e^{-100} + e^{-50 \etath^3} \right) ,
\end{equation}
we find the following fitting formulae for the functions in Equations~(\ref{eq:J_th_fitted},\ref{eq:A_th_fitted}):
\begin{equation}
   \rho(\etath)  = \kappa^{-0.5} e^{-1000/\etath^2},
\end{equation}
\begin{equation}
   \zeta(\etath)  = e^{-10^{-7}/\etath^2},
\end{equation}
\begin{equation}
   \lambda_1(\etath) =  4.43 \,\etath^{1.015}\kappa^{1.015} \left(1-e^{-80/\etath^{2.3}}\right)\times
   10^{ \frac{1 + 0.150\etath^{2/3} -0.591\etath^{4/3} + 0.195\etath^{6/3}}{1 +2.236\etath^{2/3}-3.312\etath^{4/3} + 1.10\etath^{6/3}}} + 7.435\kappa^{1.015} e^{-80/\etath^{1.5}},
\end{equation}
\begin{equation}
   \lambda_2(\etath) = 0.970 \,\etath^{1.015} \kappa^{1.015}\left(1-e^{-50/\etath^{1.5}}\right)\times 10^{\frac{1 + 20.69\etath^{2/3} -8.81\etath^{4/3} + 0.941\etath^{6/3}}{1 +33.77\etath^{2/3}-17.43\etath^{4/3} + 2.87\etath^{6/3}}} + 4.703 \kappa^{1.015}e^{-50/\etath},
\end{equation}
\begin{equation}
   \mu_1(\etath)  = 0.706 \,\etath^{1.015} \kappa^{1.015} \left(1-e^{-80/\etath^{1.5}}\right)\times10^{ \frac{1 + 2.34\times 10^5\etath^{2/3} -1.8\times10^5\etath^{4/3} + 2.32\times10^4\etath^{6/3}}{1 +1.206\times10^5\etath^{2/3}-1.01\times10^5\etath^{4/3} + 4.82\times10^4\etath^{6/3}}} + 9.4073\kappa^{1.015} e^{-80/\etath},
\end{equation}
\begin{equation}
   \mu_2(\etath)  = 0.07 \,\etath^{1.015}\kappa^{1.015} \,\, \left(1-e^{-40/\etath^{1.5}}\right)\times10^{ \frac{1 + 1.115\times 10^5\etath^{2/3} -6.18\times10^4\etath^{4/3} + 5.82\times10^4\etath^{6/3}}{1 +7.54\times10^4\etath^{2/3}+4.03\times10^4\etath^{4/3} + 9.81\times10^4\etath^{6/3}}} + 5.679\kappa^{1.015} e^{-40/\etath}.
\end{equation}
Examples of the final thermal fitting functions are shown in Figure~\ref{fig:TH_coeff} for several values of $\etath$; the mean relative error of the coefficients as a function of $\etath$ is displayed in Figure~\ref{fig:TH_error}. The maximum relative errors in the second panel have characteristic values on the order of $1\%$, with the error at the peak---where the high and low asymptotic limits are joined---rising to $10$--$25\%$ depending on the value of $\etath$. The mean relative error stays $\lesssim 10\%$ for all $\etath$. 
For $\etath \sim10^2-10^6$, the error rises for large $y$ because the approximate saddle point solution (Equation~\ref{eq:zs_fit}) does not replicate the quartic root to high-enough accuracy.\footnote{
The steepest descent approximation is exponentially sensitive to the value of $z_s$. To see this, consider the term $\exp{(-y/z^2)}$. The correct steepest descent approximation is $\exp{(-y/z_s^2)}$. If we have $z_{s, \rm fitted} = z_s + \epsilon$, the approximation picks up a relative error of $ 1 - \exp (\epsilon\, y/y_0)$. Even for $\epsilon\ll1$, then at some large enough frequency $y$ we may end up with a relative error of order unity. For large $\etath$, the value of $\epsilon$ rises as $y$ increases, leading to the error seen in Figure~\ref{fig:TH_coeff}.} 
Nevertheless, this error only occurs for large $y$ where the emission and absorption coefficients are exponentially suppressed, and thus the approximate form for $z_s$ is sufficient our purposes. Because of this small discrepancy in the value of the saddle point, the mean relative error shown in Figure~\ref{fig:TH_coeff} rises slightly, but is still under good control. For $\etath\gg1$, the fitting functions given here are accurate to within $4.88\%$ and $5.32\%$, respectively. Our fitting functions thus have an error slightly higher than the $2.7\%$ quoted by \cite{Mahadevan96} for large $\etath$, but are able to handle every value of $\etath$ with reasonable accuracy.

\section{Application to Afterglow Modeling}\label{sec:afterglow}
 \begin{figure*}
 \centering
  \includegraphics[width=0.5\textwidth]{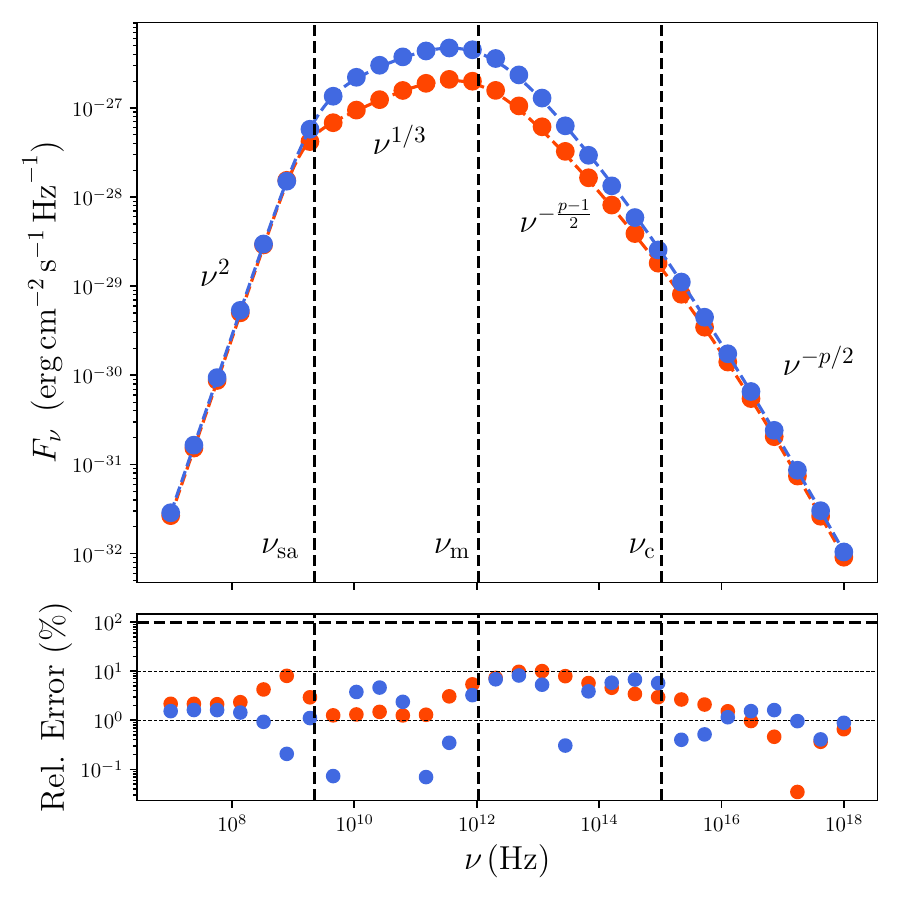}
  \caption{A test case for the power-law fitting functions derived in \S\ref{sec:pl_dist}. We insert the fitting functions into the numerical framework of \citealt{FM26} as applied to late-time gamma-ray burst afterglows and compare the results to the analytic spectrum 1 of \citealt{GS02}. The top panel compares the values of the overall specific flux for two models of the ambient number density, a constant-density interstellar medium (indicated by $k=0$) and a stellar-wind density falling as $r^{-2}$ ($k=2$). The frequency scalings of the flux and the key break frequencies (for $k=2$ only, though the $k=0$ break frequencies are nearly identical) are indicated on the top panel, similar to the top panel of Figure 1 in \citealt{GS02}. The bottom panel shows the percent relative error between each calculation. The error is always below $10\%$, with typical values $\sim 1\%$. This comparison confirms that the analytic fitting functions derived in this work can be successfully used in place of computationally expensive numerical integrals when computing emission from synchrotron afterglows.
  }
  \label{fig:GS02_comparison}
  \end{figure*}
tA primary application of the analytic cooled radiation coefficients derived in this work is to modeling synchrotron afterglows from shock waves. The cooled coefficients allow for a full examination of the impact of different modeling assumptions (e.g., hydrodynamics and mildly relativistic shock speeds) on afterglow spectra and light curves. We leave a full analysis to future work; in this section, we simply wish to furnish an example using the cooled coefficients  to analyze GRB afterglows. To this end, we assume Blandford-McKee hydrodynamics and insert Equations~(\ref{eq:J_pl_fitted}), ~(\ref{eq:A_pl_fitted}), ~(\ref{eq:J_th_fitted}), ~(\ref{eq:A_th_fitted}) into the full-volume model of \cite{FM26} (see Appendix D of that work for a discussion of the full-volume code as applied to ultrarelativistic Blandford-McKee GRB afterglows). Cooling in the Blandford-McKee context is discussed in Appendix~\ref{appendix:Blandford-McKee}. We compare this computation to the analytic spectrum calculated by \cite{GS02}, who assume a distribution consisting only of power-law electrons and numerically calculate the emission and absorption coefficients. Note that the fitting formulas derived in the present work are not intended to replace the functions derived by \cite{GS02} and used in this section. The \cite{GS02} results are applicable in calculating spectral energy distributions and light curves for on-axis, highly relativistic blast waves whose hydrodynamics follows the Blandford-McKee solution. The fitting functions derived in the present work require the use of detailed numerical models to calculate SEDs and light curves, but have the advantage of being applicable in broader contexts.

Figure~\ref{fig:GS02_comparison} demonstrates the results of each calculation for a `slow cooling' GRB spectrum (spectrum 1 in Figure 1 of \citealt{GS02}), where the characteristic cooling frequency $\nu_c$ is greater than the synchrotron frequency of the lowest-energy power-law electrons $\nu_m$. These characteristic frequencies are related to the local parameters $x_1$ and $x_2=x_{\infty}$ discussed in this work by a suitable emission-weighted average over emitting fluid elements. Locally, the minimal synchrotron frequency and cooling frequency occur, respectively, at $\nu(x_1=1) = \nu_0 \gamma_1$ and $\nu(x_\infty=1) =\nu_0 \gamma_\infty$. Denoting the emissivity-weighted average schematically using angle brackets, the observed characteristic frequencies are then $\nu_m \sim\langle\nu_0 \gamma_1^2\rangle$ and $\nu_c \sim\langle\nu_0 \gamma_\infty^2\rangle$. In this spectrum, the self-absorption frequency $\nu_{\rm sa}$ is smaller than $\nu_{m}$. The fiducial values (see \citealt{FM26} for a description of each variable) are $p=2.5$, $\epsilon_e=0.1$, $\epsilon_B=0.1$, $\epsilon_T=0.4$, $n_0 = 10^{-1} \,\rm cm^{-3}$, $z=0$, $d_L = 10^{28} \,\rm cm$, $T = 1 \,\rm day$, and $(\Gamma\beta)_{\rm sh, 0} =10$. 
The ambient number density is assumed to scale as $r^{-k}$ and the shock velocity as $R^{-\alpha}$; for the Blandford-McKee solution these parameters are related via $\alpha = (k-3)/2$. The orange points/curves correspond to $k=0$, with blue points/curves indicating $k=2$. Note that $(\Gamma\beta)_{\rm sh, 0}$ and $n_0$ as defined in Equations 11 and 12 of \cite{FM26} are the proper shock velocity and external number density at the maximum perpendicular shock radius $R_0$ (Equation 4 of \citealt{FM26}). In the ultrarelativistic limit, $R_0$ and the radius $R_l$ along the line-of-sight (LOS) can be calculated in terms of $(\Gamma\beta)_{\rm sh, 0}$ and $\alpha$ via Equations B21 and B22 of \cite{FM26}. In particular, $R_0 \simeq (2\alpha + 1) (\alpha+1)^{-1} \pvnot^2 \, cT$. Using these relations and writing the mass density as $m_p n_0  \equiv A R_0^{-k}$, the explosion energy in units of $10^{52}  \,\rm erg$ is given by Equation A15 of \cite{GS02} as

\begin{equation}
    E = 5.25\times10^{53} \rm erg\, \,\frac{(4-k)^3}{(5-k)^3 (17-4k)} \left(\frac{n_0}{1 \,\,cm^{-3}}\right) \,\left(\frac{\pvnot}{10}\right)^8 \left(\frac{T}{1 \,\,\rm day}\right)^3.
\end{equation}
Note that this formula strictly applies only in the ultrarelativistic limit. Using the slightly more exact numerical approach of \cite{FM26} for the parameters given above, the respective explosion energies for $k=0$ and $k=2$ are  $1.62\times10^{51} \,\rm erg$ and $1.76\times10^{51} \,\rm erg$.

As shown in the second panel of Figure~\ref{fig:GS02_comparison}, the analytic coefficients of the previous sections are effective in capturing the physics of synchrotron shocks. The relative error is always below $10\%$ compared to the fitting functions of \cite{GS02}, with the typical error hovering around $1\%$ (note that the analytic fitting functions of \citealt{GS02} are only accurate to within $7\%$ themselves). We conclude that the analytic fitting functions derived in this work are suitable for use in full-volume models of synchrotron-emitting shock waves.

\section{Conclusion}\label{sec:conclusion}

In this paper, we have explored the effects of cooling on synchrotron-emitting electrons. Tracking the evolution of a single advecting fluid element, the radiation coefficients are typically strongly impacted by both radiative and adiabatic cooling, which alter the shape of the emitting electron distribution \citep{GS02}. The presence of cooling leads to an exponential decrease in both the emission and absorption coefficients at high frequencies. The highest-energy electrons cool faster than lower-energy electrons, leading to a narrower distribution in energy. As time increases, the emission and absorption coefficients change their form further, and converge to Dirac delta-functions in energy. 
In general, the Dirac-delta form of the cooled distributions may occur before the bulk of the electrons becomes nonrelativistic. For example, considering a power-law distribution with negligible adiabatic cooling, $\gamma_2$ cools significantly before $\gamma_1$ begins to cool appreciably, and the Dirac-delta form of the distribution can clearly be achieved for ultrarelativistic velocities. Encapsulating all of this behavior in a single analytic formula requires complex fitting functions. Our key results are analytic solutions to the cooled distribution functions (Equations~\ref{eq:pl_dist_cooled}, \ref{eq:therm_dist_cooled}) and fitting functions for the resulting synchrotron emissivity and absorption coefficients from cooled power-law electrons (Equations~\ref{eq:J_pl_fitted}, \ref{eq:A_pl_fitted}) and thermal electrons (Equations~\ref{eq:J_th_fitted}, \ref{eq:A_th_fitted}).

 The general formalism for handling cooling is described in \S\ref{sec:distributions}, with results applicable to arbitrary impulsively injected electron distributions. In \S\ref{sec:pl_dist} and \S\ref{sec:th_dist}, we apply these results to power-law and thermal distributions with the aim of deriving fitting functions for the synchrotron radiation coefficients. The altered functional forms of the smooth-cutoff cooled electron distributions typically differ from the simpler modeling choice which cuts off the distribution at the maximal Lorentz factor (\S\ref{sec:FO14}). An analytic calculation of the cooled power-law distribution and a numerical calculation of the resulting emission and absorption coefficients has been carried out for application to GRB afterglows \citep[e.g.,][]{GS02,ResslerLaskar17,Warren22}. The analytic fitting functions provided in this work can be used to quickly and effectively replicate these results (\S\ref{sec:afterglow}).

The primary application of our fitting functions is to full-volume models of synchrotron emission from astrophysical strong shocks. Inserting the fitting functions into numerical codes allows us to consistently capture the cooling break of the combined synchrotron spectrum, as in Figure~\ref{fig:GS02_comparison}. In addition to replicating the results of \cite{GS02}, these fitting functions allow for the full-volume synchrotron-emitting shock formalism to be extended to more general contexts. To apply the local cooling coefficients, the local fluid properties of each element from the time of injection to the time of emission must be specified to calculate the functions $\mathscr{F}$ and $\mathscr{G}$. In one-zone models, cooling is typically accounted for in power-law distributions by enforcing a steeper spectral index at large Lorentz factors, which correctly reproduces the $\nu^{-p/2}$ flux scaling above the cooling break. In the full-volume local picture, this scaling instead arises as the average emission from many fluid elements with varied degrees of cooling. We note that the use of these fitting functions is not limited to shock emission, since the formalism used is applicable to synchrotron emission from any impulsively injected population of electrons cooling over time.

In deriving our results we have assumed that synchrotron emission from relativistic electrons is dominant. The solution obtained for the cooled distributions is only valid when $\gamma\gg1$; indeed, the solution allows for the unphysical case $\gamma<1$. When $\gamma\lesssim2$ (and, in the thermal case, for $\Theta\lesssim1$), we must modify Equation~(\ref{eq:cooling_ODE}) to enforce $\gamma\geq1$ and account for the presence of cyclo-synchrotron emission \citep{Mahadevan96}. In practical cases, this mildly relativistic and nonrelativistic emission is suppressed compared to synchrotron emission from ultrarelativistic electrons, and is often negligible. We caution, however, that our current results are not applicable in these regimes. For the power-law case, we have also assumed that $p>2$, such that the injected distribution is governed by the minimum Lorentz factor (for $p<2$, one must also specify a finite maximum Lorentz factor $\gamma_2$). A treatment of a power-law distributions with $p<2$ would be conceptually similar to our current analysis, but would yield qualitatively different results and is outside the scope of our present work.

The construction of the cooled thermal distributions assumes that electrons are thermalized only at the time of injection and subsequently evolve solely through cooling. In other words, it is implicitly assumed that the electrons are not heated/re-thermalized downstream from energy exchange with ions or other electrons. In the application to shock-accelerated particles, Coulomb collisions are typically too slow to effectively mediate energy exchange, and the possibility of having thermalized particles is assumed to be related to collisionless processes operating near the shock front. In these cases, then, it is reasonable to assume that no mechanism can re-thermalize electrons as they advect downstream. In any case, adding heating terms would significantly alter Equation~(\ref{eq:cooling_ODE}), potentially precluding the possibility of obtaining closed-form analytic solutions to the distribution functions---a primary focus of our work.

The emitting electrons in this work are assumed to cool via two mechanisms: radiative synchrotron cooling and adiabatic cooling. The effects of inverse-Compton cooling can trivially be added into the existing framework by modifying the synchrotron loss timescale $t_B/\gamma$, as discussed in \S\ref{sec:distributions}.
Other sources such as bremsstrahlung and Coulomb cooling can, in principle, also be included in the formalism of this work \cite[see e.g.,][]{VurmMetzger17}, but would require more significant modifications. A key limitation of the cooling formalism of Equations~(\ref{eq:distribution_evolution}, \ref{eq:cooling_ODE}) is when synchrotron self-absorption becomes important. In the self-absorbed regime, electrons can be heated by absorbing synchrotron radiation from other particles, and the downstream particle distribution must be solved for using a full kinetic equation rather than Equation~(\ref{eq:cooling_ODE}; e.g., \citealt{McCray69, Ghisellini88, Ghisellini98, Gao13}). To the best of our knowledge this effect has not been studied in the context of a full-volume model (though see Appendix~D of \citealt{Rahaman25}), and deserves further investigation.

For ease of use, the fitting functions derived in this work are available as a \texttt{Python} module on Zenodo: \dataset[synchrotron\_cooling\_fitting\_functions]{https://doi.org/10.5281/zenodo.22084984}.

\section*{acknowledgments}
B.M. and R.F. 
are supported in part by the National Science Foundation under grant number AST-2508620. R.F. thanks Jonathan Granot for helpful discussions and the Yukawa Institute for Theoretical Physics at Kyoto University, at which some of this work was done during the long-term workshop "Multi-Messenger Astrophysics in the Dynamic Universe." We would like to thank the referee for their detailed and insightful comments that helped improve the clarity of the manuscript.
\bibliography{refs}{}

@ARTICLE{Margalit&Quataert21,
       author = {{Margalit}, Ben and {Quataert}, Eliot},
        title = "{Thermal Electrons in Mildly Relativistic Synchrotron Blast Waves}",
      journal = {\apjl},
         year = 2021,
        month = dec,
       volume = {923},
       number = {1},
          eid = {L14},
        pages = {L14},
          doi = {10.3847/2041-8213/ac3d97},
archivePrefix = {arXiv},
       eprint = {2111.00012},
 primaryClass = {astro-ph.HE},
       adsurl = {https://ui.adsabs.harvard.edu/abs/2021ApJ...923L..14M}
}

@ARTICLE{Granot+99a,
       author = {{Granot}, Jonathan and {Piran}, Tsvi and {Sari}, Re'em},
        title = "{Images and Spectra from the Interior of a Relativistic Fireball}",
      journal = {\apj},
         year = 1999,
        month = mar,
       volume = {513},
       number = {2},
        pages = {679-689},
          doi = {10.1086/306884},
archivePrefix = {arXiv},
       eprint = {astro-ph/9806192},
 primaryClass = {astro-ph},
       adsurl = {https://ui.adsabs.harvard.edu/abs/1999ApJ...513..679G}
}

@ARTICLE{Granot+99b,
       author = {{Granot}, Jonathan and {Piran}, Tsvi and {Sari}, Re'em},
        title = "{Synchrotron Self-Absorption in Gamma-Ray Burst Afterglow}",
      journal = {\apj},
         year = 1999,
        month = dec,
       volume = {527},
       number = {1},
        pages = {236-246},
          doi = {10.1086/308052},
archivePrefix = {arXiv},
       eprint = {astro-ph/9808007},
 primaryClass = {astro-ph},
       adsurl = {https://ui.adsabs.harvard.edu/abs/1999ApJ...527..236G}
}

@ARTICLE{BlandfordMcKee76,
       author = {{Blandford}, R.~D. and {McKee}, C.~F.},
        title = "{Fluid dynamics of relativistic blast waves}",
      journal = {Physics of Fluids},
         year = 1976,
        month = aug,
       volume = {19},
        pages = {1130-1138},
          doi = {10.1063/1.861619},
       adsurl = {https://ui.adsabs.harvard.edu/abs/1976PhFl...19.1130B}
}

@ARTICLE{Chevalier98,
       author = {{Chevalier}, Roger A.},
        title = "{Synchrotron Self-Absorption in Radio Supernovae}",
      journal = {\apj},
         year = 1998,
        month = may,
       volume = {499},
       number = {2},
        pages = {810-819},
          doi = {10.1086/305676},
       adsurl = {https://ui.adsabs.harvard.edu/abs/1998ApJ...499..810C}
}

@ARTICLE{Bell,
       author = {{Bell}, A.~R.},
        title = "{The acceleration of cosmic rays in shock fronts - I.}",
      journal = {\mnras},
         year = 1978,
        month = jan,
       volume = {182},
        pages = {147-156},
          doi = {10.1093/mnras/182.2.147},
       adsurl = {https://ui.adsabs.harvard.edu/abs/1978MNRAS.182..147B}
}

@ARTICLE{BO78,
       author = {{Blandford}, R.~D. and {Ostriker}, J.~P.},
        title = "{Particle acceleration by astrophysical shocks.}",
      journal = {\apjl},
         year = 1978,
        month = apr,
       volume = {221},
        pages = {L29-L32},
          doi = {10.1086/182658},
       adsurl = {https://ui.adsabs.harvard.edu/abs/1978ApJ...221L..29B}
}

@ARTICLE{BE87,
       author = {{Blandford}, Roger and {Eichler}, David},
        title = "{Particle acceleration at astrophysical shocks: A theory of cosmic ray origin}",
      journal = {\physrep},
         year = 1987,
        month = oct,
       volume = {154},
       number = {1},
        pages = {1-75},
          doi = {10.1016/0370-1573(87)90134-7},
       adsurl = {https://ui.adsabs.harvard.edu/abs/1987PhR...154....1B}
}

@ARTICLE{Spitkovsky08,
       author = {{Spitkovsky}, Anatoly},
        title = "{Particle Acceleration in Relativistic Collisionless Shocks: Fermi Process at Last?}",
      journal = {\apjl},
         year = 2008,
        month = jul,
       volume = {682},
       number = {1},
        pages = {L5},
          doi = {10.1086/590248},
archivePrefix = {arXiv},
       eprint = {0802.3216},
 primaryClass = {astro-ph},
       adsurl = {https://ui.adsabs.harvard.edu/abs/2008ApJ...682L...5S}
}

@ARTICLE{Park11,
       author = {{Park}, Jaehong and {Caprioli}, Damiano and {Spitkovsky}, Anatoly},
        title = "{Simultaneous Acceleration of Protons and Electrons at Nonrelativistic Quasiparallel Collisionless Shocks}",
      journal = {\prl},
         year = 2015,
        month = feb,
       volume = {114},
       number = {8},
          eid = {085003},
        pages = {085003},
          doi = {10.1103/PhysRevLett.114.085003},
archivePrefix = {arXiv},
       eprint = {1412.0672},
 primaryClass = {astro-ph.HE},
       adsurl = {https://ui.adsabs.harvard.edu/abs/2015PhRvL.114h5003P}
}

@ARTICLE{Crumley19,
       author = {{Crumley}, P. and {Caprioli}, D. and {Markoff}, S. and {Spitkovsky}, A.},
        title = "{Kinetic simulations of mildly relativistic shocks - I. Particle acceleration in high Mach number shocks}",
      journal = {\mnras},
         year = 2019,
        month = jun,
       volume = {485},
       number = {4},
        pages = {5105-5119},
          doi = {10.1093/mnras/stz232},
archivePrefix = {arXiv},
       eprint = {1809.10809},
 primaryClass = {astro-ph.HE},
       adsurl = {https://ui.adsabs.harvard.edu/abs/2019MNRAS.485.5105C}
}

@ARTICLE{Margutti19,
       author = {{Margutti}, R. and {Metzger}, B.~D. and {Chornock}, R. and {Vurm}, I. and {Roth}, N. and {Grefenstette}, B.~W. and {Savchenko}, V. and {Cartier}, R. and {Steiner}, J.~F. and {Terreran}, G. and {Margalit}, B. and {Migliori}, G. and {Milisavljevic}, D. and {Alexander}, K.~D. and {Bietenholz}, M. and {Blanchard}, P.~K. and {Bozzo}, E. and {Brethauer}, D. and {Chilingarian}, I.~V. and {Coppejans}, D.~L. and {Ducci}, L. and {Ferrigno}, C. and {Fong}, W. and {G{\"o}tz}, D. and {Guidorzi}, C. and {Hajela}, A. and {Hurley}, K. and {Kuulkers}, E. and {Laurent}, P. and {Mereghetti}, S. and {Nicholl}, M. and {Patnaude}, D. and {Ubertini}, P. and {Banovetz}, J. and {Bartel}, N. and {Berger}, E. and {Coughlin}, E.~R. and {Eftekhari}, T. and {Frederiks}, D.~D. and {Kozlova}, A.~V. and {Laskar}, T. and {Svinkin}, D.~S. and {Drout}, M.~R. and {MacFadyen}, A. and {Paterson}, K.},
        title = "{An Embedded X-Ray Source Shines through the Aspherical AT 2018cow: Revealing the Inner Workings of the Most Luminous Fast-evolving Optical Transients}",
      journal = {\apj},
         year = 2019,
        month = feb,
       volume = {872},
       number = {1},
          eid = {18},
        pages = {18},
          doi = {10.3847/1538-4357/aafa01},
archivePrefix = {arXiv},
       eprint = {1810.10720},
 primaryClass = {astro-ph.HE},
       adsurl = {https://ui.adsabs.harvard.edu/abs/2019ApJ...872...18M}
}

@ARTICLE{Ho19,
       author = {{Ho}, Anna Y.~Q. and {Phinney}, E. Sterl and {Ravi}, Vikram and {Kulkarni}, S.~R. and {Petitpas}, Glen and {Emonts}, Bjorn and {Bhalerao}, V. and {Blundell}, Ray and {Cenko}, S. Bradley and {Dobie}, Dougal and {Howie}, Ryan and {Kamraj}, Nikita and {Kasliwal}, Mansi M. and {Murphy}, Tara and {Perley}, Daniel A. and {Sridharan}, T.~K. and {Yoon}, Ilsang},
        title = "{AT2018cow: A Luminous Millimeter Transient}",
      journal = {\apj},
         year = 2019,
        month = jan,
       volume = {871},
       number = {1},
          eid = {73},
        pages = {73},
          doi = {10.3847/1538-4357/aaf473},
archivePrefix = {arXiv},
       eprint = {1810.10880},
 primaryClass = {astro-ph.HE},
       adsurl = {https://ui.adsabs.harvard.edu/abs/2019ApJ...871...73H}
}

@ARTICLE{Burrows11,
       author = {{Burrows}, D.~N. and {Kennea}, J.~A. and {Ghisellini}, G. and {Mangano}, V. and {Zhang}, B. and {Page}, K.~L. and {Eracleous}, M. and {Romano}, P. and {Sakamoto}, T. and {Falcone}, A.~D. and {Osborne}, J.~P. and {Campana}, S. and {Beardmore}, A.~P. and {Breeveld}, A.~A. and {Chester}, M.~M. and {Corbet}, R. and {Covino}, S. and {Cummings}, J.~R. and {D'Avanzo}, P. and {D'Elia}, V. and {Esposito}, P. and {Evans}, P.~A. and {Fugazza}, D. and {Gelbord}, J.~M. and {Hiroi}, K. and {Holland}, S.~T. and {Huang}, K.~Y. and {Im}, M. and {Israel}, G. and {Jeon}, Y. and {Jeon}, Y. -B. and {Jun}, H.~D. and {Kawai}, N. and {Kim}, J.~H. and {Krimm}, H.~A. and {Marshall}, F.~E. and {P. M{\'e}sz{\'a}ros} and {Negoro}, H. and {Omodei}, N. and {Park}, W. -K. and {Perkins}, J.~S. and {Sugizaki}, M. and {Sung}, H. -I. and {Tagliaferri}, G. and {Troja}, E. and {Ueda}, Y. and {Urata}, Y. and {Usui}, R. and {Antonelli}, L.~A. and {Barthelmy}, S.~D. and {Cusumano}, G. and {Giommi}, P. and {Melandri}, A. and {Perri}, M. and {Racusin}, J.~L. and {Sbarufatti}, B. and {Siegel}, M.~H. and {Gehrels}, N.},
        title = "{Relativistic jet activity from the tidal disruption of a star by a massive black hole}",
      journal = {\nat},
         year = 2011,
        month = aug,
       volume = {476},
       number = {7361},
        pages = {421-424},
          doi = {10.1038/nature10374},
archivePrefix = {arXiv},
       eprint = {1104.4787},
 primaryClass = {astro-ph.HE},
       adsurl = {https://ui.adsabs.harvard.edu/abs/2011Natur.476..421B}
}

@ARTICLE{ResslerLaskar17,
       author = {{Ressler}, Sean M. and {Laskar}, Tanmoy},
        title = "{Thermal Electrons in Gamma-Ray Burst Afterglows}",
      journal = {\apj},
         year = 2017,
        month = aug,
       volume = {845},
       number = {2},
          eid = {150},
        pages = {150},
          doi = {10.3847/1538-4357/aa8268},
archivePrefix = {arXiv},
       eprint = {1706.01885},
 primaryClass = {astro-ph.HE},
       adsurl = {https://ui.adsabs.harvard.edu/abs/2017ApJ...845..150R}
}

@ARTICLE{Warren22,
       author = {{Warren}, Donald C. and {Dainotti}, Maria and {Barkov}, Maxim V. and {Ahlgren}, Bj{\"o}rn and {Ito}, Hirotaka and {Nagataki}, Shigehiro},
        title = "{A Semianalytic Afterglow with Thermal Electrons and Synchrotron Self-Compton Emission}",
      journal = {\apj},
         year = 2022,
        month = jan,
       volume = {924},
       number = {1},
          eid = {40},
        pages = {40},
          doi = {10.3847/1538-4357/ac2f43},
archivePrefix = {arXiv},
       eprint = {2109.07687},
 primaryClass = {astro-ph.HE},
       adsurl = {https://ui.adsabs.harvard.edu/abs/2022ApJ...924...40W}
}

@ARTICLE{GianniosSpitkovsky09,
       author = {{Giannios}, Dimitrios and {Spitkovsky}, Anatoly},
        title = "{Signatures of a Maxwellian component in shock-accelerated electrons in GRBs}",
      journal = {\mnras},
         year = 2009,
        month = nov,
       volume = {400},
       number = {1},
        pages = {330-336},
          doi = {10.1111/j.1365-2966.2009.15454.x},
archivePrefix = {arXiv},
       eprint = {0905.1970},
 primaryClass = {astro-ph.HE},
       adsurl = {https://ui.adsabs.harvard.edu/abs/2009MNRAS.400..330G}
}

@ARTICLE{Weiler86,
       author = {{Weiler}, K.~W. and {Sramek}, R.~A. and {Panagia}, N. and {van der Hulst}, J.~M. and {Salvati}, M.},
        title = "{Radio Supernovae}",
      journal = {\apj},
         year = 1986,
        month = feb,
       volume = {301},
        pages = {790},
          doi = {10.1086/163944},
       adsurl = {https://ui.adsabs.harvard.edu/abs/1986ApJ...301..790W}
}

@ARTICLE{Sari98,
       author = {{Sari}, Re'em and {Piran}, Tsvi and {Narayan}, Ramesh},
        title = "{Spectra and Light Curves of Gamma-Ray Burst Afterglows}",
      journal = {\apjl},
         year = 1998,
        month = apr,
       volume = {497},
       number = {1},
        pages = {L17-L20},
          doi = {10.1086/311269},
archivePrefix = {arXiv},
       eprint = {astro-ph/9712005},
 primaryClass = {astro-ph},
       adsurl = {https://ui.adsabs.harvard.edu/abs/1998ApJ...497L..17S}
}

@ARTICLE{GS02,
       author = {{Granot}, Jonathan and {Sari}, Re'em},
        title = "{The Shape of Spectral Breaks in Gamma-Ray Burst Afterglows}",
      journal = {\apj},
         year = 2002,
        month = apr,
       volume = {568},
       number = {2},
        pages = {820-829},
          doi = {10.1086/338966},
archivePrefix = {arXiv},
       eprint = {astro-ph/0108027},
 primaryClass = {astro-ph},
       adsurl = {https://ui.adsabs.harvard.edu/abs/2002ApJ...568..820G}
}

@ARTICLE{Ozel,
       author = {{{\"O}zel}, Feryal and {Psaltis}, Dimitrios and {Narayan}, Ramesh},
        title = "{Hybrid Thermal-Nonthermal Synchrotron Emission from Hot Accretion Flows}",
      journal = {\apj},
         year = 2000,
        month = sep,
       volume = {541},
       number = {1},
        pages = {234-249},
          doi = {10.1086/309396},
archivePrefix = {arXiv},
       eprint = {astro-ph/0004195},
 primaryClass = {astro-ph},
       adsurl = {https://ui.adsabs.harvard.edu/abs/2000ApJ...541..234O}
}

@BOOK{RBL,
       author = {{Rybicki}, George B. and {Lightman}, Alan P.},
        title = "{Radiative processes in astrophysics}",
         year = 1979,
       adsurl = {https://ui.adsabs.harvard.edu/abs/1979rpa..book.....R}
}

@ARTICLE{Nakar11,
       author = {{Nakar}, Ehud and {Piran}, Tsvi},
        title = "{Detectable radio flares following gravitational waves from mergers of binary neutron stars}",
      journal = {\nat},
         year = 2011,
        month = oct,
       volume = {478},
       number = {7367},
        pages = {82-84},
          doi = {10.1038/nature10365},
archivePrefix = {arXiv},
       eprint = {1102.1020},
 primaryClass = {astro-ph.HE},
       adsurl = {https://ui.adsabs.harvard.edu/abs/2011Natur.478...82N}
}

@ARTICLE{Berger14,
       author = {{Berger}, Edo},
        title = "{Short-Duration Gamma-Ray Bursts}",
      journal = {\araa},
         year = 2014,
        month = aug,
       volume = {52},
        pages = {43-105},
          doi = {10.1146/annurev-astro-081913-035926},
archivePrefix = {arXiv},
       eprint = {1311.2603},
 primaryClass = {astro-ph.HE},
       adsurl = {https://ui.adsabs.harvard.edu/abs/2014ARA&A..52...43B}
}

@book{python,
  author    = {Van Rossum, Guido and Drake, Fred L.},
  title     = {Python 3 Reference Manual},
  year      = {2009},
  isbn      = {1441412697},
  publisher = {CreateSpace},
  address   = {Scotts Valley, CA}
}

@ARTICLE{Kathirgamaraju2019,
       author = {{Kathirgamaraju}, Adithan and {Giannios}, Dimitrios and {Beniamini}, Paz},
        title = "{Observable features of GW170817 kilonova afterglow}",
      journal = {\mnras},
         year = 2019,
        month = aug,
       volume = {487},
       number = {3},
        pages = {3914-3921},
          doi = {10.1093/mnras/stz1564},
archivePrefix = {arXiv},
       eprint = {1901.00868},
 primaryClass = {astro-ph.HE},
       adsurl = {https://ui.adsabs.harvard.edu/abs/2019MNRAS.487.3914K}
}

@ARTICLE{Jikei25,
       author = {{Jikei}, Taiki and {Groselj}, Daniel and {Sironi}, Lorenzo},
        title = "{Magnetic Field Amplification and Particle Acceleration in Weakly Magnetized Trans-relativistic Electron-ion Shocks}",
      journal = {arXiv e-prints},
         year = 2025,
        month = dec,
          eid = {arXiv:2512.03169},
        pages = {arXiv:2512.03169},
          doi = {10.48550/arXiv.2512.03169},
archivePrefix = {arXiv},
       eprint = {2512.03169},
 primaryClass = {astro-ph.HE},
       adsurl = {https://ui.adsabs.harvard.edu/abs/2025arXiv251203169J}
}

@ARTICLE{Aharonian10,
       author = {{Aharonian}, F.~A. and {Kelner}, S.~R. and {Prosekin}, A. Yu.},
        title = "{Angular, spectral, and time distributions of highest energy protons and associated secondary gamma rays and neutrinos propagating through extragalactic magnetic and radiation fields}",
      journal = {\prd},
         year = 2010,
        month = aug,
       volume = {82},
       number = {4},
          eid = {043002},
        pages = {043002},
          doi = {10.1103/PhysRevD.82.043002},
archivePrefix = {arXiv},
       eprint = {1006.1045},
 primaryClass = {astro-ph.HE},
       adsurl = {https://ui.adsabs.harvard.edu/abs/2010PhRvD..82d3002A}
}

@ARTICLE{Crusius86,
       author = {{Crusius}, A. and {Schlickeiser}, R.},
        title = "{Synchrotron radiation in random magnetic fields}",
      journal = {\aap},
         year = 1986,
        month = aug,
       volume = {164},
       number = {2},
        pages = {L16-L18},
       adsurl = {https://ui.adsabs.harvard.edu/abs/1986A&A...164L..16C}
}

@BOOK{Zhang19,
       author = {{Zhang}, Bing},
        title = "{The physics of gamma-ray bursts}",
         year = 2019,
       adsurl = {https://ui.adsabs.harvard.edu/abs/2019pgrb.book.....Z}
}

@ARTICLE{AguilarRuiz25,
       author = {{Aguilar-Ruiz}, Edilberto and {Gill}, Ramandeep and {Beniamini}, Paz and {Granot}, Jonathan},
        title = "{Synchrotron self-compton model of TeV afterglows in gamma-ray bursts}",
      journal = {\mnras},
         year = 2026,
        month = mar,
       volume = {546},
       number = {3},
          eid = {stag101},
        pages = {stag101},
          doi = {10.1093/mnras/stag101},
archivePrefix = {arXiv},
       eprint = {2511.23349},
 primaryClass = {astro-ph.HE},
       adsurl = {https://ui.adsabs.harvard.edu/abs/2026MNRAS.546ag101A}
}

@ARTICLE{Yuan03,
       author = {{Yuan}, Feng and {Quataert}, Eliot and {Narayan}, Ramesh},
        title = "{Nonthermal Electrons in Radiatively Inefficient Accretion Flow Models of Sagittarius A*}",
      journal = {\apj},
         year = 2003,
        month = nov,
       volume = {598},
       number = {1},
        pages = {301-312},
          doi = {10.1086/378716},
archivePrefix = {arXiv},
       eprint = {astro-ph/0304125},
 primaryClass = {astro-ph},
       adsurl = {https://ui.adsabs.harvard.edu/abs/2003ApJ...598..301Y}
}

@ARTICLE{Giannios09,
       author = {{Giannios}, Dimitrios and {Spitkovsky}, Anatoly},
        title = "{Signatures of a Maxwellian component in shock-accelerated electrons in GRBs}",
      journal = {\mnras},
         year = 2009,
        month = nov,
       volume = {400},
       number = {1},
        pages = {330-336},
          doi = {10.1111/j.1365-2966.2009.15454.x},
archivePrefix = {arXiv},
       eprint = {0905.1970},
 primaryClass = {astro-ph.HE},
       adsurl = {https://ui.adsabs.harvard.edu/abs/2009MNRAS.400..330G}
}

@ARTICLE{FM26,
       author = {{Ferguson}, Ross and {Margalit}, Ben},
        title = "{Numerical Modeling of Relativistic Effects in Synchrotron-emitting Shocks}",
      journal = {\apj},
         year = 2026,
        month = mar,
       volume = {1000},
       number = {1},
          eid = {16},
        pages = {16},
          doi = {10.3847/1538-4357/ae3a93},
archivePrefix = {arXiv},
       eprint = {2509.16313},
 primaryClass = {astro-ph.HE},
       adsurl = {https://ui.adsabs.harvard.edu/abs/2026ApJ..1000...16F}
}

@ARTICLE{FO14,
       author = {{Fouka}, M. and {Ouichaoui}, S.},
        title = "{Analytical fits for the synchrotron emission from a power-law particle distribution with a sharp cutoff}",
      journal = {\mnras},
         year = 2014,
        month = aug,
       volume = {442},
       number = {2},
        pages = {979-994},
          doi = {10.1093/mnras/stu922},
       adsurl = {https://ui.adsabs.harvard.edu/abs/2014MNRAS.442..979F}
}

@BOOK{BenderOrszag78,
       author = {{Bender}, C.~M. and {Orszag}, S.~A.},
        title = "{Advanced Mathematical Methods for Scientists and Engineers}",
         year = 1978,
       adsurl = {https://ui.adsabs.harvard.edu/abs/1978amms.book.....B}
}

@ARTICLE{Mahadevan96,
       author = {{Mahadevan}, Rohan and {Narayan}, Ramesh and {Yi}, Insu},
        title = "{Harmony in Electrons: Cyclotron and Synchrotron Emission by Thermal Electrons in a Magnetic Field}",
      journal = {\apj},
         year = 1996,
        month = jul,
       volume = {465},
        pages = {327},
          doi = {10.1086/177422},
archivePrefix = {arXiv},
       eprint = {astro-ph/9601073},
 primaryClass = {astro-ph},
       adsurl = {https://ui.adsabs.harvard.edu/abs/1996ApJ...465..327M}
}

@ARTICLE{Petrosian81,
       author = {{Petrosian}, V.},
        title = "{Synchrotron emissivity from mildly relativistic particles}",
      journal = {\apj},
         year = 1981,
        month = dec,
       volume = {251},
        pages = {727-738},
          doi = {10.1086/159517},
       adsurl = {https://ui.adsabs.harvard.edu/abs/1981ApJ...251..727P}
}

@ARTICLE{Ghisellini88,
       author = {{Ghisellini}, Gabriele and {Guilbert}, Paul W. and {Svensson}, Roland},
        title = "{The Synchrotron Boiler}",
      journal = {\apjl},
         year = 1988,
        month = nov,
       volume = {334},
        pages = {L5},
          doi = {10.1086/185300},
       adsurl = {https://ui.adsabs.harvard.edu/abs/1988ApJ...334L...5G}
}

@ARTICLE{VurmMetzger17,
       author = {{Vurm}, Indrek and {Metzger}, Brian D.},
        title = "{High-energy Emission from Nonrelativistic Radiative Shocks: Application to Gamma-Ray Novae}",
      journal = {\apj},
         year = 2018,
        month = jan,
       volume = {852},
       number = {1},
          eid = {62},
        pages = {62},
          doi = {10.3847/1538-4357/aa9c4a},
archivePrefix = {arXiv},
       eprint = {1611.04532},
 primaryClass = {astro-ph.HE},
       adsurl = {https://ui.adsabs.harvard.edu/abs/2018ApJ...852...62V}
}

@ARTICLE{McCray69,
       author = {{McCray}, Richard},
        title = "{Synchrotron Radiation Losses in Self-Absorbed Radio Sources}",
      journal = {\apj},
         year = 1969,
        month = apr,
       volume = {156},
        pages = {329},
          doi = {10.1086/149968},
       adsurl = {https://ui.adsabs.harvard.edu/abs/1969ApJ...156..329M}
}

@ARTICLE{Rahaman25,
       author = {{Rahaman}, Sk. Minhajur and {Granot}, Jonathan and {Beniamini}, Paz},
        title = "{Cooling Regimes of Nonthermal Electrons: The Slow, the Fast, and the Very Fast}",
      journal = {\apjl},
         year = 2025,
        month = aug,
       volume = {988},
       number = {2},
          eid = {L68},
        pages = {L68},
          doi = {10.3847/2041-8213/aded07},
       adsurl = {https://ui.adsabs.harvard.edu/abs/2025ApJ...988L..68R}
}

@ARTICLE{Ghisellini98,
       author = {{Ghisellini}, G. and {Haardt}, F. and {Svensson}, R.},
        title = "{Thermalization by synchrotron absorption in compact sources: electron and photon distributions}",
      journal = {\mnras},
         year = 1998,
        month = jun,
       volume = {297},
       number = {2},
        pages = {348-354},
          doi = {10.1046/j.1365-8711.1998.01442.x},
archivePrefix = {arXiv},
       eprint = {astro-ph/9712166},
 primaryClass = {astro-ph},
       adsurl = {https://ui.adsabs.harvard.edu/abs/1998MNRAS.297..348G}
}

@ARTICLE{Gao13,
       author = {{Gao}, He and {Lei}, Wei-Hua and {Wu}, Xue-Feng and {Zhang}, Bing},
        title = "{Compton scattering of self-absorbed synchrotron emission}",
      journal = {\mnras},
         year = 2013,
        month = nov,
       volume = {435},
       number = {3},
        pages = {2520-2531},
          doi = {10.1093/mnras/stt1461},
archivePrefix = {arXiv},
       eprint = {1204.1386},
 primaryClass = {astro-ph.HE},
       adsurl = {https://ui.adsabs.harvard.edu/abs/2013MNRAS.435.2520G}
}

@ARTICLE{Sironi13,
       author = {{Sironi}, Lorenzo and {Spitkovsky}, Anatoly and {Arons}, Jonathan},
        title = "{The Maximum Energy of Accelerated Particles in Relativistic Collisionless Shocks}",
      journal = {\apj},
         year = 2013,
        month = jul,
       volume = {771},
       number = {1},
          eid = {54},
        pages = {54},
          doi = {10.1088/0004-637X/771/1/54},
archivePrefix = {arXiv},
       eprint = {1301.5333},
 primaryClass = {astro-ph.HE},
       adsurl = {https://ui.adsabs.harvard.edu/abs/2013ApJ...771...54S}
}

@ARTICLE{Bilbao24,
       author = {{Bilbao}, P.~J. and {Ewart}, R.~J. and {Assun{\c{c}}ao}, F. and {Silva}, T. and {Silva}, L.~O.},
        title = "{Ring momentum distributions as a general feature of Vlasov dynamics in the synchrotron dominated regime}",
      journal = {Physics of Plasmas},
         year = 2024,
        month = may,
       volume = {31},
       number = {5},
          eid = {052112},
        pages = {052112},
          doi = {10.1063/5.0206813},
archivePrefix = {arXiv},
       eprint = {2404.11586},
 primaryClass = {physics.plasm-ph},
       adsurl = {https://ui.adsabs.harvard.edu/abs/2024PhPl...31e2112B}
}

@ARTICLE{Bilbao23,
       author = {{Bilbao}, P.~J. and {Silva}, L.~O.},
        title = "{Radiation Reaction Cooling as a Source of Anisotropic Momentum Distributions with Inverted Populations}",
      journal = {\prl},
         year = 2023,
        month = apr,
       volume = {130},
       number = {16},
          eid = {165101},
        pages = {165101},
          doi = {10.1103/PhysRevLett.130.165101},
archivePrefix = {arXiv},
       eprint = {2212.12271},
 primaryClass = {physics.plasm-ph},
       adsurl = {https://ui.adsabs.harvard.edu/abs/2023PhRvL.130p5101B}
}
\bibliographystyle{aasjournal}

\appendix

\section{Cooling in Blandford-McKee Hydrodynamics}\label{appendix:Blandford-McKee}

In modeling ultra-relativistic shocks, the analytic Blandford-McKee hydrodynamic solution \citep{BlandfordMcKee76} is often used to describe the post-shock fluid. Blandford-McKee hydrodynamics is especially convenient for examining the effects of particle cooling because the functions $\mathscr{G}$ and $\mathscr{F}$, and thus the complete form of the particle distributions, may be solved analytically. For power-law distributions, this has been done previously in \citep{GS02}. We find it useful to repeat the calculation here, following their approach, to illustrate the generalized formalism of \S~\ref{sec:distributions} and to find an explicit form for the cooled thermal distribution.

Using the self-similar variable $\chi$, the relation between the local and injected ($\chi = 1)$ values of the electron number density, fluid Lorentz factor, observer time $t$, and thermal energy density $u_{\rm th} \propto B^2/8\pi$ in the Blandford-McKee solution may be written (assuming the external mass density $\rho\propto r^{-k}$) \citep[e.g.,][]{GS02}
\begin{equation}
    \frac{n_e}{n_{e, \rm inj}} = \chi^{-\frac{13-2k}{2(4-k)}}, \qquad
    \frac{\gamma}{\gamma_{ \rm inj}} = \chi^{-\frac{7-2k}{2(4-k)}}, \qquad
    \frac{t}{t_{\rm inj}} = \chi^{\frac{1}{4-k}}, \qquad
    \frac{u_{\rm th} }{u_{\rm th, inj} } = \chi^{-\frac{2(13-2k)}{3(4-k)}}.
\end{equation}
The observer time $t$ is related to the time $t'$ in the fluid rest frame via $dt' = dt/\gamma$,\footnote{The quantities in \S~\ref{sec:distributions} were evaluated in the fluid rest frame. To clarify the distinction between the observer frame and the fluid rest frame in this appendix, here we denote such quantities with a prime.} so we also have 
\begin{equation}
    dt' = \frac{t_{\rm inj}}{\gamma_{\rm inj}}\frac{1}{4-k} \chi^{\frac{1}{2(4-k)}} d\chi.
\end{equation}
Using the above relations, we can compute the cooling timescales introduced in \S~\ref{sec:distributions}:
\begin{equation}
    t_{\rm ad}'^{-1} = \frac{13-2k}{6}\frac{\gamma_{\rm inj}}{t_{\rm inj}} \chi^{-\frac{9-2k}{2(4-k)}}, \qquad
    t_B'^{-1} = \frac{\sigma_T}{6\pi m_ec} B_{\rm inj}^2 \,\chi^{-\frac{2(13-2k)}{3(4-k)}}.
\end{equation}
We consider three cases: radiative and adiabatic cooling, radiative cooling only ($t'_{\rm ad}=\infty$), and adiabatic cooling only ($t'_{\rm B}=\infty$). The values of $\mathscr{G}(\chi)$ and $\mathscr{F}(\chi)$ in each case are
\begin{flalign}
    \mathscr{G}(\chi) & = \exp\left[\displaystyle\int dt' \,\,t_{\rm ad}'^{-1}\right]  = \exp\left[\frac{13-2k}{6(4-k)}\displaystyle\int_1^\chi d\chi' \,\chi'^{-1}\right] & \nonumber \\
    & = \begin{cases} 
       \chi^{\frac{13-2k}{6(4-k)}}& \rm rad.+ad. \\
      1 & \rm rad.\, only\\         
       \chi^{\frac{13-2k}{6(4-k)}}& \rm ad.\, only\\ \end{cases},     
\end{flalign}
\begin{flalign}
    \mathscr{F}(\chi) & = 
    \int dt'_1 \,t_B'^{-1}\exp\left[- \displaystyle\int_{t_{\rm inj}'}^{t_1'} dt_2 \,\,t'^{-1}_{\rm ad}\right] = \frac{\sigma_T}{6\pi m_ec} B_{\rm inj}^2 \int_1^{\chi} d\chi' \,\chi'^{-\frac{2(13-2k)}{3(4-k)}} \mathscr{G}^{-1}(\chi) & \nonumber \\
    & = \begin{cases} 
      \frac{\sigma_T}{2(19-2k)\pi m_ec} \frac{B_{\rm inj}^2 t_{\rm inj}}{\gamma_{\rm inj}} \left[1-\chi^{-\frac{19-2k}{3(4-k)}}\right] & \rm rad.+ad. \\
      \frac{\sigma_T}{(25-2k)\pi m_ec} \frac{B_{\rm inj}^2 t_{\rm inj}}{\gamma_{\rm inj}} \left[1-\chi^{-\frac{25-2k}{6(4-k)}}\right] & \rm rad.\, only\\         
      0& \rm ad.\, only\\ \end{cases}.
\end{flalign}
Using these formulas,
\begin{flalign}
    \gamma_{\rm \infty}(\chi) & = 
    \frac{1}{\mathscr{G}(\chi)\mathscr{F}(\chi)}& \nonumber \\
    & = \begin{cases} 
      \frac{2(19-2k)\pi m_ec}{\sigma_T} \frac{\gamma_{\rm inj}}{B_{\rm inj}^2 t_{\rm inj}} \frac{\chi^{\frac{25-2k}{6(4-k)}}}{\chi^{\frac{19-2k}{3(4-k)}}-1} & \rm rad.+ad. \\
           \frac{(25-2k)\pi m_ec}{\sigma_T} \frac{\gamma_{\rm inj}}{B_{\rm inj}^2 t_{\rm inj}} \frac{\chi^{\frac{25-2k}{6(4-k)}}}{\chi^{\frac{25-2k}{6(4-k)}}-1} & \rm rad.\, only\\         
      \infty& \rm ad.\, only\\ \end{cases}.
\end{flalign}
The first line is in agreement with the calculation presented in \cite{GS02} (their Equation A12). Finally, using Equations~(\ref{eq:pl_dist_cooled},\ref{eq:therm_dist_cooled}), the cooled distributions may be written as a function of the self-similar coordinate $\chi$ behind the shock (assuming both radiative and adiabatic cooling)
\begin{equation}\label{eq:pl_dist_cooled_BM}
    \left(\frac{\partial n_e}{\partial \gamma}\right)_{\rm pl}^{\rm BM} = K_{\rm inj}\,\chi^{-\frac{(p+2)(13-2k)}{6(4-k)}}\,\,
    \gamma^{-p} \left(1- \frac{\gamma}{\gamma_{\infty}(\chi)}\right)^{p-2},
\end{equation}
\begin{equation}\label{eq:therm_dist_cooled_BM}
    \left(\frac{\partial n_e}{\partial \gamma}\right)_{\rm th}^{\rm BM} = L_{\rm inj} \,\frac{\gamma^2}{2\Theta_{\rm inj}^3}\, \frac{1}{[ 1- \gamma/\gamma_{\infty}(\chi) ]^4}\,\exp \left[ -\frac{\chi^{\frac{(13-2k)}{6(4-k)}}}{\Theta_{\rm inj}} \frac{\gamma}{1- \gamma/\gamma_{\infty}(\chi)}\right]  \times\sqrt{1-\chi^{-\frac{2(13-2k)}{6(4-k)}}\frac{[1- \gamma/\gamma_{\infty}(\chi)]^2}{\gamma^2}}.
\end{equation}

\section{Pitch-Angle Averaged Synchrotron Function}\label{appendix:synchrotron_function}

For describing synchrotron emission in a randomly oriented magnetic field, it is useful to define a ``pitch-angle averaged" synchrotron function as follows. Denoting a pitch-angle averaged quantity $X$ as $\tilde{X}$, we average over the single-particle power to obtain \citep{ResslerLaskar17}
\begin{equation}\label{eq:pitch_angle_power}
    \tilde{P}_e = \int_0^{\pi/2} d\alpha\,\, P_e(\sin{\alpha}) = \frac{\sqrt{3}e^3 B}{m_e c^2}\int_0^{\pi/2} d\alpha\,\, \sin^2{\alpha} \,\,F(x/\sin{\alpha}) = \frac{\sqrt{3}e^3 B}{m_e c^2}\tilde{F}(x),
\end{equation}
where $x = 4\pi m_e c\nu/3eB\gamma^2$, $\alpha$ is the electron's pitch-angle, and the synchrotron function is defined in terms of a modified Bessel function by $F(x) = x \int_x^\infty dy\,\,K_{5/3}(y)$. The pitch-angle averaging can thus be absorbed into the function $\tilde{F}(x)$.

For $x\ll 1$, $F(x) \to F_0 x^{1/3}$, where $F_0 = 2^{5/3}\pi/\sqrt{3}\,\Gamma(1/3)$ \citep{RBL}. Thus, the pitch-angle-averaged synchrotron function satisfies
\begin{flalign}
&& \tilde{F}(x\ll1) & = F_0\int_0^{\pi/2} d\alpha \,\, \sin^{5/3}{\alpha} \,\, x^{1/3}  & \nonumber \\
&& & = \frac{\sqrt{3} \,2^{5/3} \pi^{3/2}\,\Gamma(4/3)}{5\,\,\Gamma(1/3)\,\Gamma(5/6)} x^{1/3}  & \nonumber \\
&& & \equiv \tilde{F}_0\, x^{1/3}.
\end{flalign}
For $x\gg 1$, we have instead $F(x)\to\sqrt{\pi/2}\,e^{-x} x^{1/2}$. Then,
\begin{equation}
\tilde{F}(x\gg1)  = \sqrt{\frac{\pi}{2}} \,x^{1/2}\int_0^{\pi/2} d\alpha \,\, \sin^{3/2}{\alpha} \,\,  e^{-x/\sin{\alpha}}.
\end{equation}
Using the integral on the right-hand side may be approximated as $\sqrt{\pi/2x} \,\,e^{-x}$ using the method of steepest descent. Plugging this in above, we get the simple expression
\begin{equation}
\tilde{F}(x\gg1)  = \frac{\pi}{2} e^{-x}.
\end{equation}
The function $\tilde{F}(x)$ may be written exactly in terms of Whittaker functions \citep{Crusius86} or modified Bessel functions \citep{Aharonian10}, but it is more convenient to use the approximate form given in Eq. D7 of \cite{Aharonian10}
\footnote{Note the change in labels for $F(x)$ and $\tilde{F}(x)$ in this work versus \cite{Aharonian10}.},
\begin{equation}\label{eq:F_tilde}
\tilde{F}(x) = \frac{1.808}{\sqrt{1+3.4x^{2/3}}} \frac{1 + 2.21x^{2/3} + 0.347 x^{4/3}}{1 + 1.353x^{2/3} + 0.217 x^{4/3}} \,\, x^{1/3}e^{-x}.
\end{equation}
Similarly, we use the approximate form given in \cite{Aharonian10} for the synchrotron function $F(x)$, ignoring pitch-angle effects ($\sin{\alpha}=1$):
\begin{equation}
F(x) = 2.15 (1 + 3.06x) ^{1/6} \frac{1 + 0.884x^{2/3} + 0.471 x^{4/3}}{1 + 1.64x^{2/3} + 0.974 x^{4/3}} \,\, x^{1/3}e^{-x}.
\end{equation}
For the calculation of absorption coefficients, it is useful to define another function $H(x)$ as
\begin{equation}\label{eq:H}
H(x) \equiv -\frac{d }{dx}\left(\frac{F(x)}{x}\right) = K_{5/3}(x).
\end{equation}
The non-pitch-angle averaged $H(x)$ has the leading-order asymptotic limits
\begin{equation}
H(x) = \begin{cases} 
      \frac{2}{3} F_0 x^{-5/3} & x\ll1 \\
      \sqrt{\frac{\pi}{2}} x^{-1/2}  e^{-x} & x\gg1
   \end{cases} ,
\end{equation}
Replacing $F$ with $\tilde{F}$ in Equation~(\ref{eq:H}), the pitch-angle averaged version satisfies
\begin{equation}\label{eq:H_tilde_limits}
\tilde{H}(x) = \begin{cases} 
      \frac{2 }{3} \tilde{F}_0 x^{-5/3} & x\ll1 \\
      \frac{\pi}{2} x^{-1}  e^{-x} & x\gg1
   \end{cases}.
\end{equation}
Suitable fitting functions for $H$ and $\tilde{H}$ (accurate to within $0.71\%$ and $3.69\%$, respectively) are
\begin{equation}
H(x) =  1.433(1+0.764 x^{0.89})^{1.31} \frac{1 + 0.181 x^{2/3} + 0.628 x^{4/3}}{1 + 0.33 x^{2/3} + 0.505^{4/3}} x^{-5/3} e^{-x}.
\end{equation}
\begin{equation}\label{eq:H_tilde}
\tilde{H}(x) =  1.206(1+1.45 x^{1.262})^{0.53} \frac{1 - 0.4615 x^{2/3} + 2.35 x^{4/3}}{1 - 0.656 x^{2/3} + 2.192 x^{4/3}} x^{-5/3} e^{-x}.
\end{equation}
\section{Fitting Functions Without Pitch-Angle Averaging}\label{app:perp_pitch_angle}

In certain cases, it is advantageous to consider synchrotron emission from cooled electrons in the presence of a strong background magnetic field. For such applications, we provide analogous fitting functions to those derived in the main text under the assumption that the distribution of pitch-angles is perpendicular to the magnetic field direction, $\sin \alpha = 1$. The methods used to derive these fitting functions mirror those used in the main text, with the synchrotron function replacements $\tilde{F}$ to $F$ and $\tilde{H}$ to $H$. The same saddle point approximations as those in the main text may be used in the perpendicular pitch-angle case. To distinguish the present functions from those derived in the main text, we add a ``$\perp$" superscript. The errors present in these functions are similar to those of their pitch-angle-averaged counterparts. Corrections have been added in several places to minimize error in the $\eta-1\sim 1$ and $\etath\sim1$ regimes.

\subsection{Power-Law Functions}

For perpendicular pitch-angles, the power-law fitting functions take the forms
\begin{equation}\label{eq:J_pl_perp_fitted}
    J_{\rm pl}^\perp(p, \, x_1, x_\infty) = \Omega_p^\perp(\eta, x_\infty) S_1^\perp(p,\eta,x_1) + \Psi_p^\perp(x_\infty) S_2^\perp(p,\eta,x_1),
\end{equation}
\begin{equation}\label{eq:A_pl_perp_fitted}
    A_{\rm pl}^\perp(p, \, \eta, x_1) = \chi_p^\perp(\eta, x_\infty) S_3^\perp(p,\eta,x_1) + \Sigma_p^\perp(x_\infty) S_4^\perp(p,\eta,x_1).
\end{equation}
The low-frequency limits $\Omega_p^\perp$ and $\chi_p^\perp$ are taken to have the same form as their pitch-angle-averaged counterparts (Equations~\ref{eq:Omega},\ref{eq:chi}) with the replacements $\tilde{F} \to F$ and $\tilde{H} \to H$ (see Appendix~\ref{appendix:synchrotron_function}). The high-frequency limits are 
 \begin{equation}\label{eq:Psi_p_perp}
    \Psi_p^\perp(x_\infty) = A_1^\perp(p)\,\exp\left[-a_1^\perp x_\infty^2  - a_2^\perp x_\infty^{2/3}\right] + \sqrt{\frac{2}{\pi}}\,\psi_p\left(\frac{p-2}{2}, x_\infty\right) \,\left(1-\exp\{-a_4^\perp x_\infty^2\}\right)^{a_3^\perp},
\end{equation}
 \begin{equation}\label{eq:Sigma_p_perp}
    \Sigma_p^\perp(x_\infty) = B_1^\perp(p)  \,\exp\left[-b_1^\perp x_\infty^2 -b_2^\perp x_\infty - b_3^\perp x_\infty^{2/3}\right] +\sigma_p^\perp(x_\infty) \,\left(1-\exp[b_5^\perp \,x_\infty]\right)^{b_4^\perp},
\end{equation}
where the low-frequency limits are 
 \begin{equation}\label{eq:A1_perp}
    A_1^\perp(p) =\frac{2^{\frac{p+1}{2}}}{p+1}\Gamma\left(\frac{p}{4} + \frac{19}{12}\right)\Gamma\left(\frac{p}{4} - \frac{1}{12}\right).
\end{equation}
 \begin{equation}\label{eq:B1_perp}
    B_1^\perp(p) = 2^{p/2} \,\,\Gamma\left(\frac{p}{4} + \frac{11}{6}\right)\Gamma\left(\frac{p}{4} + \frac{1}{6}\right).
\end{equation}
and the high-frequency limit for $\Sigma_p^\perp$ is 
\begin{equation}\label{eq:chi_p_simplified_perp}
\sigma_p^\perp(x_\infty) = \sqrt{\frac{2}{\pi}} \,\,\psi_p\left(\frac{p+1}{2},x_\infty\right) + \sqrt{\frac{2}{\pi}} \,\, \psi_p
  \left(\frac{p-1}{2},x_\infty\right).
 \end{equation}
\begin{table*}
\centering
% \noindent\makebox[\textwidth]{
\begin{tabular}{l|c|c|c|c|c|c}
     Constant & $\aleph_0$ & $\aleph_1$ & $\aleph_2$ & $\aleph_3$ & $\aleph_4$ & $\aleph_5$\\
     \hline $a_1^\perp$ & \num{2.802}& \num{-2.64}&  \num{0.925} & \num{-0.142} & \num{0.008} & 0\\
     $a_2^\perp$ & \num{-5.595}& \num{5.92}&  \num{-1.75} & \num{0.248} & \num{-0.013} & 0\\
     $a_3^\perp$ & \num{26.84}& \num{ -26.93}&  \num{10.58} & \num{ -1.675} & \num{0.096} & 0\\
     $a_4^\perp$ & \num{3.423}& \num{-3.316}&  \num{1.183} & \num{ -0.184} & \num{0.010} & 0\\  
    $b_1^\perp$ & \num{0.168}& \num{-0.184}&  \num{0.072} & \num{-0.012} & \num{0.0007}& 0\\
     $b_2^\perp$ & \num{-2.99}& \num{4.737}&  \num{-1.97} & \num{0.317} & \num{-0.018}& 0 \\
     $b_3^\perp$ & \num{2.75}& \num{ -4.60}&  \num{2.247} & \num{ -0.371} & \num{0.021} & 0\\
     $b_4^\perp$ & \num{-34.95}& \num{33.93}&  \num{-10.46} & \num{ 1.36} & \num{-0.063} & 0\\
     $b_5^\perp$ & \num{925.57}& \num{-1209.69}&  \num{616.93} & \num{ -153.04} & \num{18.503}  & \num{-0.875}\\
\end{tabular}%}
\caption{Fitting constants used to define $\Psi_p^\perp$ and  $\Sigma_p^\perp$ (Equations~\ref{eq:Psi_p_perp},\ref{eq:Sigma_p_perp}). The constants are fitted to polynomials in $p$; that is, $a_i^\perp = \sum_{j=0}^{5} \aleph_j p^j$ and $b_i^\perp = \sum_{j=0}^{5} \aleph_j p^j$.}
\label{table:perp_ab_vals}
\end{table*}
The constants $a_i^\perp$ and  $b_i^\perp$ are fitted to polynomials in $p$; the results are displayed in Table~\ref{table:perp_ab_vals}. The sigmoids $S_1^\perp$, $S_2^\perp$, $S_3^\perp$, and $S_4^\perp$ are defined as
\begin{equation}\label{eq:S_1_perp}
    S_1^\perp(p,\eta,x_1) =  \exp\left[-\alpha_1^\perp x_1^{\alpha_2^\perp} \exp\left(-\frac{\alpha_3^\perp}{(\eta^2-1)^{0.8}}\right)\right]
\end{equation}
\begin{equation}\label{eq:S_2_perp}
    S_2^\perp(p,\eta,x_1) = \left[1-S_1^\perp(p,\eta,x_1)\right]^{\alpha_4^\perp}
\end{equation}
\begin{equation}\label{eq:S_1_A_perp}
    S_3^\perp(p,\eta,x_1) =  \exp\left[-\beta_1^\perp x_1^{\beta_2^\perp} \exp\left(-\frac{1}{(\eta^2-1)^{\beta_3^\perp}}\right)\right]
\end{equation}
\begin{equation}\label{eq:S_2_A_perp}
    S_4^\perp(p,\eta,x_1) = \left[1-S_3^\perp(p,\eta,x_1)\right]^{\beta_4^\perp}
\end{equation}
with constants satisfying
\begin{equation}\label{eq:alpha_1_perp}
    \alpha_1^\perp =  903.4 - 902.2(p-0.941)^{4.63\times 10^{-4}},
\end{equation}
\begin{equation}\label{eq:alpha_2_perp}
    \alpha_2^\perp = 0.622   + 0.347 p^{2/3}  -0.017 p^{4/3},
\end{equation}
\begin{equation}\label{eq:alpha_4_perp}
    \alpha_3^\perp = 1 + (0.1p-0.71)e^{-10(\eta^2-1.1)^2}.
\end{equation}
\begin{equation}\label{eq:alpha_3_perp}
    \alpha_4^\perp =   2 - 0.5e^{-(p-2)^2} + (0.538p - 1.23 + 0.5e^{-(p-2)^2}) e^{-0.01(\eta^2-1.5)^2}
\end{equation}

\begin{equation}\label{eq:beta_1_perp}
    \beta_1^\perp =  0.124 - 22.65 (p+6.1)^{-3} + 0.514 (p - 0.344) ^{-5} - 1.34\times 10^{-4} (p - 2.01)^{2.73},
\end{equation}
\begin{equation}\label{eq:beta_2_perp}
    \beta_2^\perp = 2,
\end{equation}
\begin{equation}\label{eq:beta_4_perp}
    \beta_3^\perp = 0.47,
\end{equation}
\begin{equation}\label{eq:beta_3_perp}
    \beta_4^\perp =   2.97p-3.13.
\end{equation}

\subsection{Thermal Functions}

The perpendicular-pitch-angle thermal fitting functions are
\begin{equation}
    J_{\rm th}^\perp(y,\etath) = \Pi_J^\perp(y,\etath) e^{-\lambda_1^\perp(\etath)\frac{y}{y_t} \,\zeta_{J, \perp}(\etath)} + \left[ 1 + \rho_1^\perp(\etath) \left(\frac{y}{y_t}\right)^{-0.4}\right] \sqrt{\frac{\pi}{2}}\, y^{1/2}\, \xi_{1}(y,\etath) \left[ 1- e^{-\lambda_2^\perp(\etath)\frac{y}{y_t} \,\zeta_{J, \perp}(\etath)}\right],
\end{equation}
\begin{equation}
    A_{\rm th}^{\perp}(y,\etath) = \Pi^\perp_A(y,\etath) e^{-\mu_1^\perp(\etath)\frac{y}{y_t} \,\zeta_{A, \perp}(\etath)} + \left[ 1 + \rho_2^\perp (\etath)\left(\frac{y}{y_t}\right)^{-0.5}\right] \sqrt{\frac{\pi}{2}}\, y^{-1/2}\xi_{-2}(y,\etath) \left[ 1- e^{-\mu_2^\perp(\etath)\frac{y}{y_t} \,\zeta_{A, \perp}(\etath)}\right],
\end{equation}
where $\xi_{q}(y, \etath)$ is defined as in Equation~(\ref{eq:xi}) and the low-frequency functions $\Pi_J^\perp$ and $\Pi_A^\perp$ have the general forms as Equations~(\ref{eq:pi_J}), ~(\ref{eq:pi_A}) with the pitch-angled averaged synchrotron functions replaced by their non-pitch-angle averaged counterparts. The fitting constants are defined by
\begin{equation}
   \zeta_{J, \perp}(\etath)  = e^{-10^{-7}/\etath^{1.8}},
\end{equation}
\begin{equation}
   \zeta_{A, \perp}(\etath)  = e^{-10^{-7}/\etath^2},
\end{equation}
\begin{equation}
   \rho_1^\perp(\etath)  = 0.4\etath^{0.8} (1 - e^{-120/\etath^{2.65}})  + 2.86e^{-120/\etath^2},
\end{equation}
\begin{equation}
   \rho_2^\perp(\etath)  = 3.833 - 3.83299 e^{-(\etath-1)^2},
\end{equation}
\begin{equation}
   \lambda_1^\perp(\etath) =  3.451\,\etath^{1.015} \left(1-e^{-120/\etath^{1.4}}\right)\times
   10^{ \frac{1 + 2.15\times10^7\etath^{2/3} -1.21\times10^7\etath^{4/3} + 2.03\times10^6\etath^{6/3}}{1 +1.851\times10^7\etath^{2/3}-1.975\times10^7\etath^{4/3} + 6.31\times10^6\etath^{6/3}}} + 5.0258 e^{-120/\etath^{1.2}},
\end{equation}
\begin{equation}
   \lambda_2^\perp(\etath) = 3.327 \,\etath^{1.015} \left(1-e^{-120/\etath^{1.4}}\right)\times  10^{ \frac{1 + 3.67\times10^7\etath^{2/3} -2.025\times10^7\etath^{4/3} + 3.243\times10^6\etath^{6/3}}{1 +2.96\times10^7\etath^{2/3}-3.07\times10^7\etath^{4/3} + 9.91\times10^6\etath^{6/3}}} + 4.839 e^{-120/\etath^{1.2}},
\end{equation}
\begin{equation}
   \mu_1^\perp(\etath)  = 0.205 \,\etath^{0.591}  \left(1-e^{-20/\etath^{1.2}}\right)\times10^{ \frac{1 -1.644\etath^{2/3} + 0.794\etath^{4/3} - 0.011\etath^{6/3}}{1 -0.8296\etath^{2/3}+0.316\etath^{4/3} + 0.041\etath^{6/3}}} + 1.86 e^{-20/\etath},
\end{equation}
\begin{equation}
   \mu_2^\perp(\etath)  = 0.987 \,\etath^{0.00275} \,\, \left(1-e^{-20/\etath^{1.2}}\right)\times10^{ \frac{1 + 4.08\times 10^4\etath^{2/3} -1.754\times10^4\etath^{4/3} + 2.29\times10^3\etath^{6/3}}{1 +186\etath^{2/3}-90.9\etath^{4/3} + 12.99\etath^{6/3}}} + 0.812 e^{-20/\etath}.
\end{equation}
To improve performance around $\etath=1$, it is convenient to make a correction $\mu_2 \to \mu_2 \times (1  + 4e^{-(\etath-2)^2})$.

\end{document}